%% file: main.tex
\documentclass{article}
\usepackage{arxiv}
\usepackage[utf8]{inputenc} 
\usepackage[T1]{fontenc}    
\usepackage{hyperref}       
\usepackage{url}            
\usepackage{booktabs}       
\usepackage{amsfonts}       
\usepackage{nicefrac}       
\usepackage{microtype}      
\usepackage{lipsum}
\usepackage{float}
\usepackage{graphicx}
\graphicspath{ {./images/} }
\usepackage{subcaption}  
\usepackage{float}        
\usepackage{amssymb}
\usepackage[export]{adjustbox}
\usepackage{xcolor}
\usepackage{tcolorbox}
\usepackage{appendix}           
\usepackage{amsmath}
\usepackage{enumitem}
\usepackage{framed}
\usepackage{tocloft}
\usepackage{multirow}
\usepackage{amsthm}
\usepackage{tikz}
\usepackage{caption} 
\usepackage{array}
\usepackage{tabularx}
\usepackage{booktabs}   
\usepackage{makecell}   
\usepackage{booktabs}   
\usepackage{siunitx}    
\usepackage{adjustbox}  
\usepackage{silence}
\usepackage{cite}
\usepackage{longtable}
\usepackage{tcolorbox}
\usepackage{cite}
\usepackage{setspace}
\definecolor{rulecolor}{rgb}{0.95, 0.95, 0.95}
\definecolor{ruleborder}{rgb}{0.7, 0.7, 0.7}

\usepackage[normalem]{ulem}
\definecolor{addblue}{RGB}{0,0,180}
\definecolor{removeRed}{RGB}{200,0,0}

\newenvironment{goldenrule}{%
  \MakeFramed{\advance\hsize-\width \FrameRestore}%
  \vspace{4pt}%
}{%
  \vspace{4pt}%
  \endMakeFramed%
}

\renewcommand{\arraystretch}{1.4}
\usepackage[table]{xcolor} 
\definecolor{highlight}{RGB}{255,220,220} 

\usetikzlibrary{shapes.geometric, arrows.meta, positioning} 

\usepackage{xcolor}
\usepackage{tikz}

\definecolor{removeRed}{RGB}{190,40,40}

\definecolor{cAcad}{HTML}{0072B2}\definecolor{cGov}{HTML}{882255}
\definecolor{cNews}{HTML}{D55E00}\definecolor{cNGO}{HTML}{009E73}
\definecolor{cCorp}{HTML}{E69F00}\definecolor{cOpin}{HTML}{56B4E9}
\definecolor{cRef}{HTML}{CC79A7}\definecolor{cUGC}{HTML}{9A8700}
\newcommand{\sw}[1]{\textcolor{#1}{\rule{1.6ex}{1.6ex}}}
\tcbuselibrary{skins,breakable}
\providecommand{\swbox}{\rule{1.6ex}{1.6ex}}

\title{Epistemic Substitution: How Grokipedia's AI-Generated Encyclopedia Restructures Authority}

\author{
  Aliakbar Mehdizadeh \\
  Department of Communication \\
  University of California, Davis \\
  \texttt{amehdizadeh@ucdavis.edu}
  \And Faezeh Khazaee \\
  Department of Computer Science \\
  University of California, Davis \\
  \texttt{fkhazaee@ucdavis.edu}
  \And
  Martin Hilbert \\
  Department of Communication \\
  University of California, Davis \\
  \texttt{hilbert@ucdavis.edu}
}

\begin{document}
\maketitle

\begin{abstract}
For more than two decades, Wikipedia has represented a decentralized, human-curated model of encyclopedic knowledge production. The emergence of AI-generated encyclopedias, such as Grokipedia, raises a fundamental question: do human- and AI-curated knowledge systems rely on the same foundations of epistemic authority? We examine this question through a comparative analysis of 1,000 matched Wikipedia--Grokipedia article pairs, comprising nearly 600,000 citation instances. Using an eight-category classification scheme, we construct the epistemic profiles of articles to examine the institutional composition of their cited sources. Our findings reveal substantial differences in sourcing practices. Grokipedia relies proportionally less on academic and journalistic sources and more on user-generated content and NGO and think-tank publications. These differences persist across broad topic domains, although their magnitude varies by subject. We also find that Grokipedia adds citations approximately linearly as articles grow, whereas Wikipedia exhibits greater variation in citation density. Together, these findings suggest a shift in the institutional sources invoked to justify encyclopedic claims, which we characterize as epistemic substitution. For the same topics, the two encyclopedias rest on different institutional sources. Our findings underscore the importance of systematic, ongoing audits of information provenance for accountability in generative knowledge systems.
\end{abstract}

\section{Introduction}
\label{sec:intro}

Every society develops institutional mechanisms for determining which claims are accepted as knowledge and whose authority legitimizes them. These mechanisms have undergone successive transformations: from credentialed experts operating under institutional authority, to distributed communities negotiating consensus publicly, and now to generative AI systems that produce knowledge claims and select supporting sources without direct human involvement. Each transition relocates epistemic authority rather than merely changing who does the writing, and each therefore changes what it means for an encyclopedic claim to be justified. Encyclopedias make these authority structures particularly visible through the sources they cite to substantiate their claims. This study compares two such systems, one human-curated and one AI-generated, to examine how their citation practices reflect distinct structures of epistemic authority.

\subsection*{Theoretical Evolution of Epistemological Shifts}

To assess whether AI-generated encyclopedias represent a shift in epistemic authority, we situate their emergence within the information science tradition of examining how knowledge is organized, legitimized, and validated~\cite{hjorland2002domain}. Since its launch in 2001, Wikipedia has emerged as the dominant model of large-scale, digitally native knowledge curation~\cite{reagle2010good,mesgari2015wikipedia}. Its success demonstrated that epistemic authority could emerge through decentralized, collaborative processes rather than relying exclusively on centralized editorial oversight by credentialed experts~\cite{fallis2008toward}, as exemplified by the traditional Encyclopedia Britannica model~\cite{giles2005special}. 

Wikipedia's success rests on a human-centered model of open collaboration, consensus-building, and a policy framework that emphasizes verifiability through citations to published sources~\cite{luyt2010wikipedia}. Two core content policies and one guideline underpin this framework. \emph{No original research} (WP:NOR) prohibits unpublished original analysis or synthesis; \emph{Verifiability} (WP:V) requires that information be attributable to reliable, published sources; and the \emph{Reliable sources} guideline (WP:RS) sets out how to judge which sources qualify~\cite{wikipedia_NOR,wikipedia_V,wikipedia_RS}. Together, these rules establish a publicly documented and contestable framework for grounding encyclopedic claims in existing sources. Wikipedia's citation profile thus provides an observable, albeit imperfect, trace of its institutional standards for epistemic justification. AI-generated encyclopedias, by contrast, may produce claims through automated synthesis and select supporting sources according to processes that are not necessarily subject to equivalent public scrutiny or community-based editorial oversight.

Wikipedia has evolved from an experimental model of collaborative knowledge production into a central reference point in the digital information ecosystem, helping reshape established practices of knowledge production and validation~\cite{benkler2015peer,fallis2008toward,wray2009epistemic}. Importantly, automation has long played a role in its development. On English Wikipedia, bot-generated edits increased from 2--4\% in 2006 to about 16\% by 2009~\cite{geiger2010work}, with measurable implications for the platform's operation and governance~\cite{hilbert2020large,tsvetkova2017even}. These bots primarily perform predefined tasks, such as reverting vandalism, enforcing formatting conventions, and maintaining links, within a framework of human-defined policies and community oversight. Although some bots also generate content or modify references, their operations are generally more narrowly specified and constrained than those of contemporary generative AI systems. The key distinction, therefore, is not the presence or absence of automation, but the extent to which automated systems participate in composing encyclopedic claims and selecting the sources used to substantiate them. This distinction makes the comparison between Wikipedia and Grokipedia one of knowledge-production regimes rather than simply degrees of automation.

Traditional accounts of propositional knowledge center on epistemic subjects whose beliefs must be supported by adequate justification~\cite{gettier1963justified}. Wikipedia's distributed model of knowledge production preserves a degree of human accountability through publicly accessible revision histories, which document editorial contributions and make them open to scrutiny, even when contributors remain anonymous. LLM-based encyclopedias, by contrast, generate claims through computational processes rather than the intentional assertions of individually identifiable authors, complicating traditional notions of testimonial responsibility. This introduces a shift toward \emph{algorithmic authority}, in which computational systems increasingly mediate the production, selection, and presentation of knowledge claims~\cite{carlson2018automating}. More broadly, digital platforms are not neutral intermediaries but epistemic infrastructures whose technical architectures, governance policies, and selection mechanisms shape what information becomes visible and credible~\cite{gillespie2018custodians,van2018platform}. Research on LLM-based multi-agent systems further demonstrates that algorithmically mediated interactions can produce systematic biases, including homophily-induced network structures and emergent conformity effects~\cite{mehdizadeh2025homophily,mehdizadeh2025your}. These findings raise broader questions about how generative knowledge systems select, prioritize, and legitimize the sources used to substantiate their claims.

This transition raises fundamental questions about epistemic accountability, justification, agency, and information quality~\cite{stvilia2008information,shin2025automating,russo2024connecting}. Can generative systems meaningfully be said to assert claims when assertion traditionally presupposes a speaker who can be held accountable? Who bears responsibility for the knowledge claims these systems produce? And on what grounds can such claims be considered justified when both their content and supporting sources are selected through automated processes? These questions suggest that the emergence of LLM-based encyclopedias represents more than a technological transformation: it challenges established mechanisms through which knowledge is produced, justified, and legitimized in digital societies.

We therefore propose looking beyond encyclopedic text and its derived characteristics, such as political orientation, to examine the epistemic foundations of the sources invoked to justify its claims. Previous scholarship on Wikipedia's epistemology has often focused on \emph{epistemic consequences}, assessing the reliability, accuracy, and truthfulness of collaboratively produced knowledge~\cite{fallis2008toward}. Such studies primarily examine whether Wikipedia's distributed model of knowledge production yields reliable information. In contrast, we shift the analytical focus toward \emph{epistemic antecedents}: the sources and institutions invoked to substantiate encyclopedic claims. Following Tollefsen's account of Wikipedia as a web of testimony~\cite{tollefsen2009wikipedia}, we regard citations as observable traces of epistemic authority. When a human editor cites a scholarly journal, news outlet, or government institution, the citation constitutes an act of epistemic deference, invoking the cited source as an external authority in support of a claim. In doing so, the editor draws upon the credibility of the source while retaining responsibility for its appropriate selection and interpretation. In AI-generated encyclopedias, where automated systems participate in both producing claims and selecting supporting sources, the provenance of these citations becomes particularly important for evaluating the foundations of epistemic justification. We capture this provenance with what we call the \emph{epistemic profile}.

We define the \emph{Epistemic Profile} (EP) of an encyclopedia article as the distribution of its citations across institutional source categories, such as academic, governmental, and journalistic sources. Drawing on Buckland's concept of \emph{information-as-thing}, the EP treats citations as observable artifacts through which patterns of institutional authority can be examined~\cite{buckland1991information}. It builds on information science research on \emph{cognitive authority}: Wilson examined how people rely on second-hand knowledge and attribute authority to sources whose claims they cannot verify themselves~\cite{wilson1983second}, and Rieh showed that judgments of information quality and authority online depend on source characteristics and credibility cues~\cite{rieh2002judgment}. The EP thus makes visible which forms of institutional authority an encyclopedia invokes, and gives a measurable basis for comparing human-curated and AI-generated knowledge systems.

\subsection*{The AI Turn and the Case of Grokipedia}

In October 2025, xAI launched Grokipedia, which Elon Musk presented as an alternative to a Wikipedia he described as politically biased~\cite{grokipedia2025launch,afp2025grokipedia}. Grokipedia is built on Grok, the family of large language models that xAI has developed since 2023 and deploys as a chatbot on the social media platform X. Grok generates and revises Grokipedia's articles and, in doing so, assembles the citations that accompany them. While earlier versions included articles reproduced or adapted from Wikipedia~\cite{mohammadi2026selective}, more recent versions appear to exhibit greater autonomy in content generation, with less direct reliance on Wikipedia's text. Unlike Wikipedia, readers cannot directly edit articles but may suggest corrections for review. Content production and editorial decisions are therefore mediated by an AI system rather than governed through open, community-based editing. Beyond its stated aim of correcting political bias, Grokipedia represents a broader shift in how encyclopedic knowledge is produced, supported, and validated. Wikipedia and Grokipedia thus offer contrasting models of encyclopedic knowledge production: one grounded in human collaboration and editorial consensus, the other increasingly reliant on automated generation and source selection.

The relationship between Wikipedia and generative AI is not entirely new. Wikipedia contributed approximately 3 billion tokens to GPT-3's training data, accounting for 3\% of its weighted training mix~\cite{brown2020language}. This illustrates the reliance of early LLMs on Wikipedia, although it does not establish the composition of Grok's undisclosed training data. Yet generative AI introduces a distinction beyond the automation of existing editorial tasks: it can produce encyclopedic content and assemble supporting citations with limited direct human involvement~\cite{bommasani2021opportunities}. The question, then, is not simply whether AI-generated encyclopedias differ from human-curated ones in their textual content or political orientation, but whether they rely on the same sources to justify their claims. When Wikipedia and Grokipedia cover the same topic, do they invoke similar forms of epistemic authority, or do their citation practices reveal different epistemic profiles? We address this question through three research questions:

\begin{tcolorbox}[
  enhanced, breakable=false,
  colback=gray!5, colframe=gray!55, boxrule=0.4pt, arc=0pt,
  leftrule=2.5pt, left=8pt, right=8pt, top=6pt, bottom=6pt,
  fonttitle=\bfseries\small, coltitle=black, colbacktitle=gray!18,
  title={Research questions regarding the differences between Wikipedia and Grokipedia}]
\small
\begin{description}[font=\normalfont\bfseries, labelwidth=2.6em, leftmargin=3.2em, labelsep=0.6em, align=left, itemsep=0.6em, topsep=2pt]
  \item[RQ1] How do Wikipedia and Grokipedia differ in their use of external citations? We examine \textit{citation density} and its relationship with article length.
  \item[RQ2] Which institutions does each encyclopedia empower to warrant a claim? We operationalize this as the \textit{institutional composition} of the citation base.
  \item[RQ3] Is that structure of authority stable across subject matter, or reconfigured by topic? We operationalize this as the \textit{topic dependence} of composition, divergence, and network structure.
\end{description}
\end{tcolorbox}

\subsection*{Literature Review}

Methodologically, our study builds on research on algorithm audits~\cite{sandvig2014auditing,hilbert20248} and the evaluation of citations and information sources~\cite{greenberg2009citation,de2009bibliometrics}. Recent frameworks, such as VeriCite~\cite{qian2025vericite} and CiteEval~\cite{xu2025citeeval}, examine whether AI-generated claims are adequately supported by their citations. However, a citation can accurately support a claim while representing a different kind of epistemic authority. For example, a peer-reviewed study, a government report, and a news article may substantiate the same claim while drawing on different forms of expertise and institutional credibility. Our focus is therefore not on whether citations correctly support individual claims, but on the composition of the sources invoked to justify them.

This approach also draws on information science research on information quality and knowledge organization. Stvilia and colleagues developed frameworks for assessing information quality and examined how Wikipedia's quality assurance practices are shaped by collaborative processes~\cite{stvilia2007framework,stvilia2008information}. Hjørland emphasized that information description and classification are shaped by theoretical perspectives and interests~\cite{hjorland2023description}. Research has also demonstrated that Wikipedia's citation practices influence the dissemination of scientific knowledge, particularly by increasing the visibility of accessible academic research~\cite{teplitskiy2017amplifying}. Together, these studies highlight why the sources cited by an encyclopedia matter beyond the accuracy of its individual claims.

Recent research has examined how Grokipedia differs from Wikipedia in content, framing, and sourcing. Mohammadi and Yasseri~\cite{mohammadi2026selective} compared a large corpus of matched articles and found that about a third of Grokipedia entries closely resemble their Wikipedia counterparts, while the other two thirds are substantially rewritten; rewritten entries are longer, cite fewer references per word and, in topics such as religion, history, and literature and art, cite news outlets further to the right. Triedman and Mantzarlis~\cite{triedman2025did} examined citation reliability and the type of each platform’s 100 most-cited domains, finding more Grokipedia citations to sources rated ‘generally unreliable’ on Wikipedia’s perennial-sources list, and more academic, government and user-generated domains among Grokipedia’s most-cited hosts. Eibl et al.~\cite{eibl2026grokipedia} investigated political framing and found that semantic similarity between the platforms declines across article sections, particularly for controversial topics. Similarly, Vlahos et al.~\cite{vlahos2026grokipedia} used LLM-based evaluations of biographies of politicians and reported greater measured ideological bias in Grokipedia.

Other studies have examined structural differences and interactions between the platforms. Hadad et al.~\cite{hadad2026wikipedia} found that Grokipedia's selection and rewriting of Wikipedia articles are non-uniform, with reference density among the factors associated with rewriting. Their results also suggest that narrative structures are often preserved despite changes in content and framing. Their analysis covers the platform's launch version, collected between late October and early November 2025, whereas our corpus was collected in September 2026. Leung and Strumpf~\cite{leung2026ai}, using Grokipedia's launch as a natural experiment across 1.4 million Wikipedia biographies, found that pages with a Grokipedia counterpart subsequently received more page views and more human editing, with larger increases in editing where the Grokipedia entry was less similar to the Wikipedia article. These findings suggest that AI-generated and human-curated encyclopedias may influence one another rather than operate as entirely independent knowledge systems.

Together, these studies provide valuable evidence about textual similarity, political framing, citation reliability, and the interaction between Wikipedia and Grokipedia. However, relatively little attention has been paid to the institutional composition of full citation bases, classified citation by citation rather than by host domain. A source may be politically neutral or considered reliable while representing a different form of authority from an academic publication, government agency, or user-generated platform. We address this distinction by comparing the \emph{epistemic profiles} of matched Wikipedia and Grokipedia articles, classifying their citations into eight source categories and examining how these distributions vary across topics.

We refer to differences in the composition of cited authority between the two encyclopedias as \emph{epistemic substitution}. The term describes differences in the types of sources invoked to justify claims about the same topics, rather than the literal replacement of individual citations or a reconstruction of the processes through which claims were generated. The three research questions above measure this substitution along three dimensions: citation density, institutional composition, and topic dependence.

\section{Methodology}
\label{sec:methods}

Our comparative analysis was conducted in three phases: data collection, source classification, and network construction and analysis. 

\subsection*{Data Collection}

To construct a comparative dataset of highly edited topics, we drew candidate articles from the English Wikipedia \emph{Special:MostRevisions} report, which ranks articles by cumulative revision count, retrieved on September 2026 (Appendix~\ref{sec:appendix_data}). Each candidate was matched by title to its Grokipedia counterpart, allowing us to compare articles covering the same topic across both platforms. We excluded articles without a substantive counterpart on either platform, as well as pages primarily consisting of lists (e.g., \emph{Deaths in 2021}, \emph{List of...}). This process yielded a final corpus of 1{,}000 matched article pairs. We collected both versions of each article and extracted their citations. Further details on the scraping architecture and parsing procedures are provided in Appendix~\ref{sec:appendix_data}.

This sampling strategy intentionally focuses on highly revised articles rather than a random sample of either encyclopedia. It offers two advantages for our study. First, it provides a common set of prominent topics covered by both platforms, enabling direct comparisons of their citation practices. Second, frequently revised articles tend to receive greater editorial attention, making them particularly relevant for examining sourcing practices and institutional authority. However, revision frequency does not necessarily correspond to readership or controversy, and the resulting sample is not representative of either encyclopedia as a whole. Our findings should therefore be interpreted as characterizing differences in citation practices within this matched, high-revision subset, rather than across the entire Wikipedia or Grokipedia corpus.

\subsection{Epistemic Classification of Citations}

To examine the epistemic foundations of cited references, we developed a classification scheme detailed in the \texttt{Citation Content Coding Manual} (Appendix~\ref{subsec:coding_manual}). Rather than adopting an existing taxonomy, we designed the scheme to capture the types of sources invoked to support encyclopedic claims. This approach builds on Wilson's concept of cognitive authority and Rieh's research on credibility judgments~\cite{wilson1983second,rieh2002judgment}. Four large language models independently proposed taxonomies based on the observed sources. The authors compared these proposals and retained eight distinct categories rather than imposing a predetermined number. Each citation was classified using its available bibliographic information and context, rather than its URL alone. The development process and differences between the proposed taxonomies are documented in Appendix~\ref{subsec:coding_manual}.

Each citation was assigned to one of eight mutually exclusive categories based on the nature and provenance of the cited source (Table~\ref{tab:epistemic_categories}).

\begin{table}[htbp]
\centering
\small
\caption{\textbf{Epistemic classification scheme.} Eight mutually exclusive categories used to classify every citation.}
\label{tab:epistemic_categories}
\renewcommand{\arraystretch}{1.25}
\setlength{\tabcolsep}{5pt}
\begin{tabularx}{\textwidth}{@{} c l >{\raggedright\arraybackslash}X >{\raggedright\arraybackslash}p{0.27\textwidth} @{}}
\toprule
 & \textbf{Category} & \textbf{Definition and scope} & \textbf{Typical sources} \\
\midrule
\sw{cAcad} & Academic \& Scholarly & Peer-reviewed original research and scholarly communication. & Journal articles, conference papers, academic books \\
\sw{cGov}  & Government \& Official & Information, data, services, or legal documents produced by government bodies. & Census data, legislation, court records \\
\sw{cNews} & News \& Journalism & Original, factual reporting of current events under professional journalistic standards. & Newspaper and wire reports, broadcast news \\
\sw{cNGO}  & NGO \& Think Tank & Mission-driven research, policy papers, and advocacy reports from non-profit and civil-society organizations. & Policy briefs, NGO reports \\
\sw{cCorp} & Corporate \& Commercial & Publications by for-profit firms about their own business. & Press releases, annual reports, white papers \\
\sw{cOpin} & Opinion \& Advocacy & Persuasive commentary intended to promote a specific ideological position. & Op-eds, editorials, advocacy journalism \\
\sw{cRef}  & Reference \& Tertiary & Works that organize and index general knowledge; subject to the \textit{Look-Through Principle}. & Encyclopedias, dictionaries \\
\sw{cUGC}  & User-Generated (UGC) & Informal content from private individuals acting in a non-institutional capacity. & Tweets, forum comments, personal blogs \\
\bottomrule
\end{tabularx}
\end{table}

Importantly, these categories describe source types rather than source quality. They do not measure factual accuracy, reliability, political bias, or whether a citation adequately supports the associated claim. The classification should therefore not be interpreted as a ranking of epistemic credibility.

During pilot coding, we identified two recurring ambiguities. First, the hosting domain did not always reflect the nature of the cited work. For example, a CNN news report hosted on \texttt{youtube.com} should be classified as \textit{News \& Journalism}, rather than \textit{User-Generated Content}. Second, some citations identified specific works, such as books, while linking to pages that host or describe them, such as Google Books or the Internet Archive. Because the classification considers the full citation information rather than URLs alone, we established two decision rules (Box 1) to resolve these ambiguities. These rules were incorporated into the coding manual. The full procedure is documented in Appendix~\ref{subsec:coding_manual}.

\begin{tcolorbox}[
  enhanced, breakable=false,
  colback=gray!5, colframe=gray!55, boxrule=0.4pt, arc=0pt,
  leftrule=2.5pt, left=8pt, right=8pt, top=6pt, bottom=6pt,
  fonttitle=\bfseries\small, coltitle=black, colbacktitle=gray!18,
  title={Box 1 \,|\, Decision rules for ambiguous citations}]
\small
\begin{description}[font=\normalfont\bfseries, style=nextline, leftmargin=0pt, itemsep=0.8em, topsep=2pt]

  \item[Rule 1. Specific Work Principle]
  The classification of the \textit{specific cited work} overrides the classification of its parent domain.\\[2pt]
  {\color{black!70}\textit{Example.} An op-ed on \texttt{nytimes.com} is coded \textit{Opinion \& Advocacy}, although the domain is broadly \textit{News \& Journalism}.}

  \item[Rule 2. Look-Through Principle]
  When a citation identifies a specific work but links to a page that hosts or describes it, such as a catalog entry, a digitized copy, or a reference page, classification follows the identified work rather than the linked page. This distinction is determined using the citation's bibliographic information and context.\\[2pt]
  {\color{black!70}\textit{Example.} The Wikipedia article on Joseph Stalin cites Khlevniuk's \textit{Stalin: New Biography of a Dictator} through a link to its scan on the Internet Archive. The citation is classified by the book, as \textit{Academic \& Scholarly}, not by the archive that hosts it. Likewise, the same article cites Kotkin's \textit{Stalin: Paradoxes of Power, 1878--1928} with page numbers and links the book's title to its Wikipedia page; the author, title, and page numbers identify the book, so the citation is coded \textit{Academic \& Scholarly}, not \textit{Reference \& Tertiary}. Citations to the content of a Wikipedia article itself remain \textit{Reference \& Tertiary}.}

\end{description}
\end{tcolorbox}

Both rules apply to parts of the corpus. Under Rule~1, citations to six major news sites (\textit{The New York Times}, \textit{The Guardian}, \textit{The Washington Post}, \textit{The Wall Street Journal}, CNN, and Fox News) were classified by the type of piece: of the 584 whose URL points to an opinion section, 517 (89\%) were coded \textit{Opinion \& Advocacy} rather than \textit{News \& Journalism}. Likewise, citations to \texttt{youtube.com} were classified by who produced the video rather than by the platform: more than half were coded \textit{User-Generated Content}, and the rest mainly \textit{Corporate \& Commercial} or \textit{News \& Journalism}. Under Rule~2, about 10\% of Wikipedia citations link to books on Google Books or the Internet Archive; these were classified by the book itself, most often as \textit{Academic \& Scholarly}. Books without a link were classified from their bibliographic text in the same way.

We evaluated the reliability of automated classification in two steps. First, we compared the primary classifier (\texttt{Gemini~Flash~3.6}), which labeled the full corpus, with a second model (\texttt{Claude~Opus~5}) on a sample of 400 citations, stratified at 50 per category. The models agreed on 81.8\% of citations (Cohen's $\kappa = 0.79$). Second, two human coders evaluated 200 of these 400 citations: 150 on which the two models agreed (Stratum A) and 50 on which they disagreed (Stratum B) (Table~\ref{tab:validation_agreement}). In Stratum A ($n=150$), where the models agreed, human inter-coder agreement was 90\% ($\kappa=0.89$, 95\% CI [0.82, 0.93]), and the primary classifier matched the human consensus on 134 of the 135 citations where both coders agreed (99.3\%). In Stratum B ($n=50$), where the models disagreed, human agreement was lower at 30\% ($\kappa=0.23$, 95\% CI [0.10, 0.36]), and the primary classifier matched the human consensus on 3 of the 15 citations where both coders agreed (20\%).

These results indicate that the classifiers perform consistently on citations where the models agree, while model disagreements frequently coincide with human coding difficulties. However, the substantial human disagreement in Stratum B limits our ability to establish a definitive accuracy estimate for the full corpus. We therefore interpret distinctions between potentially overlapping categories, particularly \textit{Opinion \& Advocacy} and \textit{News \& Journalism}, with caution. Additional validation results, including per-category precision, recall, F1 scores, and examples of classification disagreements, are reported in Appendix~\ref{sec:appendix_validation}.

\begin{table}[htbp]
\centering
\small
\caption{\textbf{Agreement on the 200-citation validation sample.}
Two human coders evaluated a subset of citations. Stratum A contains citations where the models agreed ($n=150$), while Stratum B contains citations where they disagreed ($n=50$). Agreement coefficients are Cohen's $\kappa$ with 95\% bootstrap confidence intervals, except where Krippendorff's $\alpha$ is reported.}
\label{tab:validation_agreement}

\setlength{\tabcolsep}{6pt}
\renewcommand{\arraystretch}{1.15}

\begin{tabular}{@{} l r l r l @{}}
\toprule
 & \multicolumn{2}{c}{\textbf{Stratum A} ($n=150$)}
 & \multicolumn{2}{c}{\textbf{Stratum B} ($n=50$)} \\
\cmidrule(lr){2-3}\cmidrule(l){4-5}

\textbf{Comparison}
 & \% Agree & Coefficient [95\% CI]
 & \% Agree & Coefficient [95\% CI] \\
\midrule

\multicolumn{5}{@{}l}{\textit{Human--human agreement}} \\
\quad Coder A vs Coder B
 & 90 & 0.89 [0.82, 0.93]
 & 30 & 0.23 [0.10, 0.36] \\

\quad Krippendorff's $\alpha$
 & -- & 0.89
 & -- & 0.19 \\

\addlinespace

\multicolumn{5}{@{}l}{\textit{Human--model agreement}} \\

\quad Coder A vs Gemini Flash 3.6
 & 93 & 0.92 [0.87, 0.97]
 & 22 & 0.12 [0.00, 0.24] \\

\quad Coder B vs Gemini Flash 3.6
 & 95 & 0.95 [0.90, 0.98]
 & 48 & 0.38 [0.23, 0.54] \\

\quad Coder A vs Claude Opus 5
 & 93 & 0.92 [0.88, 0.97]
 & 54 & 0.46 [0.31, 0.62] \\

\quad Coder B vs Claude Opus 5
 & 95 & 0.95 [0.91, 0.98]
 & 38 & 0.29 [0.16, 0.43] \\

\bottomrule
\end{tabular}

\vspace{3pt}
\begin{minipage}{0.92\linewidth}
\footnotesize
\end{minipage}

\end{table}

\subsection{Corpus Topic Classification}

In addition to classifying citations, we categorized the 1{,}000 matched article pairs by topic to characterize the corpus and examine whether sourcing patterns vary across subject areas. We therefore applied the Wikimedia topic classifier~\cite{johnson2021language}, which assigns relevance scores across 64 topics organized into four broad domains: \textit{Culture}, \textit{Geography}, \textit{History \& Society}, and \textit{STEM}~\cite{baigutanova2023longitudinal,ruprechter2023poor,das2024language}. The classifier uses links between Wikipedia articles to predict their relevance to predefined topics. Each matched pair was assigned to the domain corresponding to its Wikipedia article's highest-scoring topic label.

This classification provides a practical way to examine sourcing differences across broad subject areas. Because Wikipedia and Grokipedia articles are matched by topic, differences in their citation profiles cannot be attributed simply to differences in the overall topic composition of the two samples. However, the broad categories do not capture all variation in source availability or relevance within individual topics.

\begin{table}[htbp]
\centering
\small
\caption{\textbf{Topic composition of the corpus.} Each of the $N=1000$ matched articles was assigned to the top-level domain of its highest-scoring label from the Wikimedia language-agnostic topic classifier; every article received a label. The last column lists representative subtopics of each domain in the Wikimedia taxonomy.}
\label{tab:corpus_topics}
\renewcommand{\arraystretch}{1.2}
\setlength{\tabcolsep}{6pt}
\begin{tabularx}{\textwidth}{@{} l r r >{\raggedright\arraybackslash}X @{}}
\toprule
\textbf{Domain} & \textbf{Articles} & \textbf{Share (\%)} & \textbf{Includes, e.g.} \\
\midrule
Culture            & 517 & 51.7 & Biography, media and entertainment, music, sports, literature, philosophy and religion \\
Geography          & 296 & 29.6 & Countries, cities, and regions of Africa, the Americas, Asia, Europe, and Oceania \\
History \& Society & 114 & 11.4 & History, politics and government, military and warfare, business and economics, education \\
STEM               & 73  & 7.3  & Physics, biology, medicine and health, technology, computing, space, earth and environment \\
\midrule
Total              & 1{,}000 & 100.0 & \\
\bottomrule
\end{tabularx}
\end{table}

\subsection*{Uses and Roles of Generative AI}

Generative AI played distinct roles in this study. First, Grokipedia, as an AI-generated encyclopedia, was itself an object of investigation. Second, generative AI assisted in developing the citation classification scheme. Four models (\texttt{ChatGPT-5}, \texttt{Claude Sonnet 4.5}, \texttt{Gemini 2.5 Pro}, and \texttt{Grok 4}) independently proposed taxonomies, which the authors reviewed and consolidated into the final eight-category scheme. Third, generative AI was used to classify citations at scale. The primary classifier, \texttt{Gemini Flash 3.6}, labeled the full corpus, while \texttt{Claude Opus 5}, an advanced thinking model, served as a secondary classifier for comparison. Classification reliability was evaluated through inter-model agreement and independent human coding. Throughout the study, the authors defined the research questions, developed the coding manual, established the validation protocol, and retained responsibility for methodological decisions and interpretation.

\section{Results}

We organize the results around our three research questions. First, we examine citation density and its relationship with article length (RQ1). Next, we compare the institutional composition of cited sources (RQ2). Finally, we investigate how this composition varies across subject areas (RQ3).

\subsection{RQ1: Citation Density and Scaling}

Table~\ref{tab:domain_stats} presents descriptive statistics for the 1{,}000 matched article pairs. Word counts were calculated from the extracted text without extensive preprocessing to remove non-lexical elements, such as tables, markup, or formatting artifacts. They should therefore be interpreted as approximate measures of article length. The unit of analysis is the citation instance: one source cited in one footnote. A Wikipedia footnote that cites several sources is split into one record per source; Grokipedia footnotes cite one source each (Appendix~\ref{sec:appendix_data}).

Grokipedia articles are, on average, longer than their Wikipedia counterparts but contain fewer citations, resulting in lower citation density (21.8 versus 34.0 citations per 1{,}000 words). This is consistent with the findings of Mohammadi and Yasseri~\cite{mohammadi2026selective}. Citation density also varies less across Grokipedia articles, with a standard deviation of 5.5 compared with 16.0 for Wikipedia. These differences suggest that Grokipedia provides fewer explicit references relative to article length, although citation density alone does not establish whether individual claims are adequately supported or factually accurate.

\begin{table}[htbp]
\centering
\small
\caption[Descriptive statistics for the article corpus]{\textbf{Article-level descriptive statistics.} Summary statistics for the $N=1{,}000$ matched article pairs. Values are reported as mean (standard deviation).}
\label{tab:domain_stats}
\renewcommand{\arraystretch}{1.25}
\setlength{\tabcolsep}{10pt}
\begin{tabular}{@{} l r r @{}}
\toprule
\textbf{Per article} & \textbf{Wikipedia} & \textbf{Grokipedia} \\
\midrule
Word count
  & 10{,}200.8 (3{,}837.0)
  & 11{,}785.4 (4{,}300.1) \\
Citation count
  & 336.7 (167.3)
  & 257.5 (112.1) \\
Citation density (per 1{,}000 words)
  & 34.0 (16.0)
  & 21.8 (5.5) \\
\bottomrule
\end{tabular}
\end{table}

\begin{figure}[htbp]
  \centering
  \includegraphics[width=\textwidth]{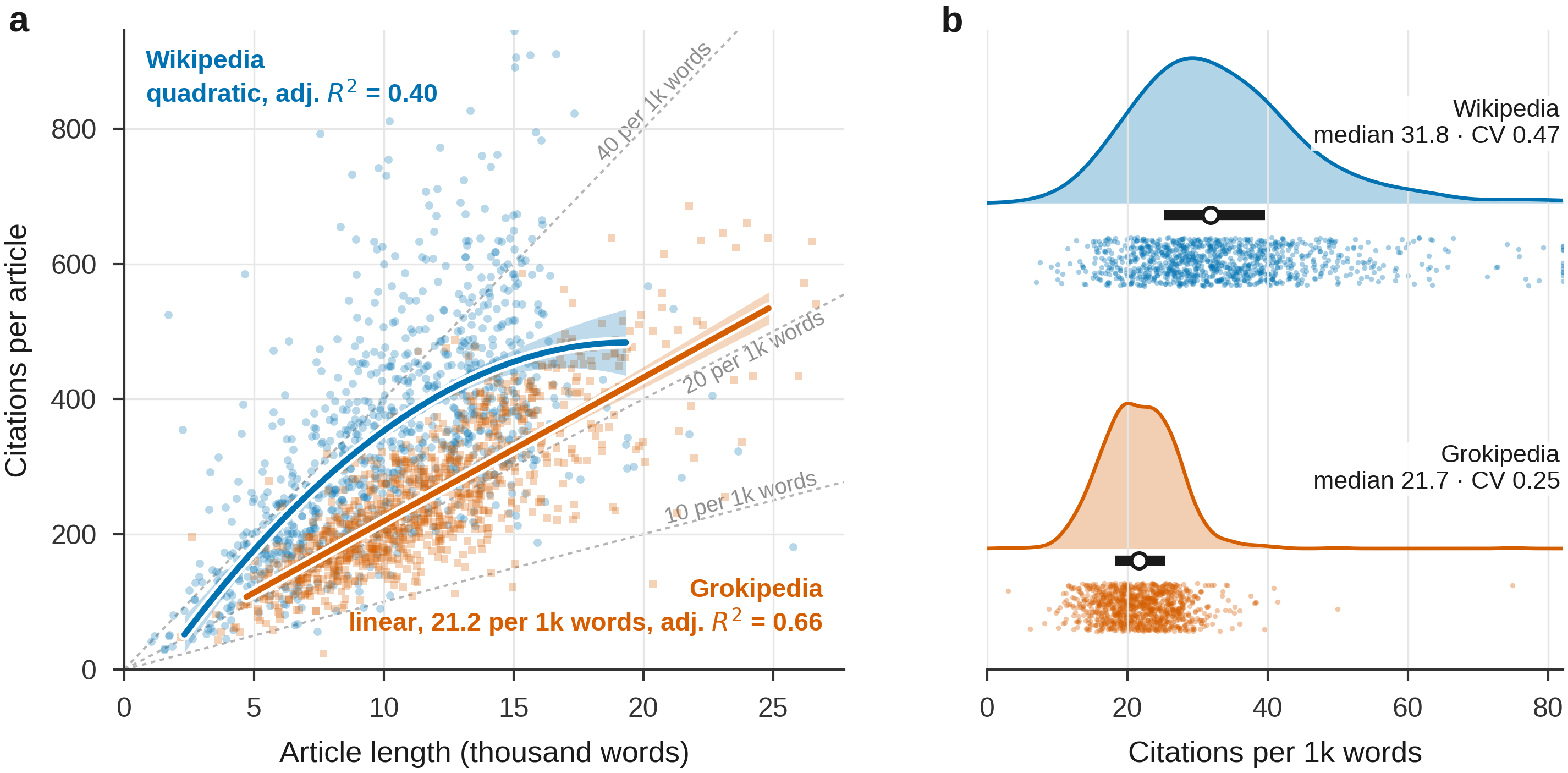}
  \caption{\textbf{Citation volume and density across article lengths.}
  \textbf{(a)} Word count against citation count; each point is one article. Lines show the selected OLS model for each platform with its 95\% confidence band: quadratic for Wikipedia (adjusted $R^2=0.40$) and linear for Grokipedia (approximately 21.2 additional citations per 1{,}000 words; adjusted $R^2=0.66$). Dashed grey lines mark constant densities of 10, 20 and 40 citations per 1{,}000 words.
  \textbf{(b)} Citation density across articles: density curves, individual articles (points), interquartile ranges (bars) and medians (white dots; Wikipedia 31.8, Grokipedia 21.7 citations per 1{,}000 words). CV: coefficient of variation.}
  \label{fig:citations_vs_words}
\end{figure}

Figure~\ref{fig:citations_vs_words} examines how citation counts scale with article length. We compared linear and quadratic ordinary least squares (OLS) models for each platform, retaining the quadratic term when statistically significant at $\alpha=0.05$. For Grokipedia, the quadratic term was not significant ($p=0.070$), and the linear model explained approximately two-thirds of the variance (adjusted $R^2=0.66$). The estimated slope was $\beta=0.021$ citations per word ($p<0.001$), corresponding to approximately 21 additional citations per 1{,}000 words. Thus, citation counts in Grokipedia increase approximately linearly with article length over the observed range.

Wikipedia exhibits a different pattern. Its quadratic term was significant ($p<0.001$), with a modest improvement in model fit over the linear specification (adjusted $R^2=0.40$ versus $0.37$). Citation counts initially increase with article length, but their growth slows for longer articles. Consistently, on a log--log scale, Grokipedia's citation count grows in proportion to article length (elasticity 1.02, 95\% CI [0.97, 1.08]), whereas Wikipedia's grows less than proportionally (0.89 [0.82, 0.96]).

Wikipedia also exhibits considerably greater variation around the fitted relationship. Overall, Grokipedia shows a more uniform relationship between article length and citation volume, whereas Wikipedia's citation practices vary more substantially across articles. Full regression results are provided in Appendix~\ref{sec:appendix_scaling}.

We also examined articles with the highest and lowest citation counts on each platform (Appendix~\ref{sec:appendix_topbottom}). Among the most heavily cited articles, Wikipedia features biographies, whereas Grokipedia features countries and regions. Notable high-citation articles include \textit{History of the United States} and \textit{Elon Musk} on Grokipedia, compared with \textit{Vladimir Lenin} and \textit{Joseph Stalin} on Wikipedia.

\subsection{RQ2: Institutional Composition of Sources}

\begin{figure}[ht!]
  \centering
  \includegraphics[width=\textwidth]{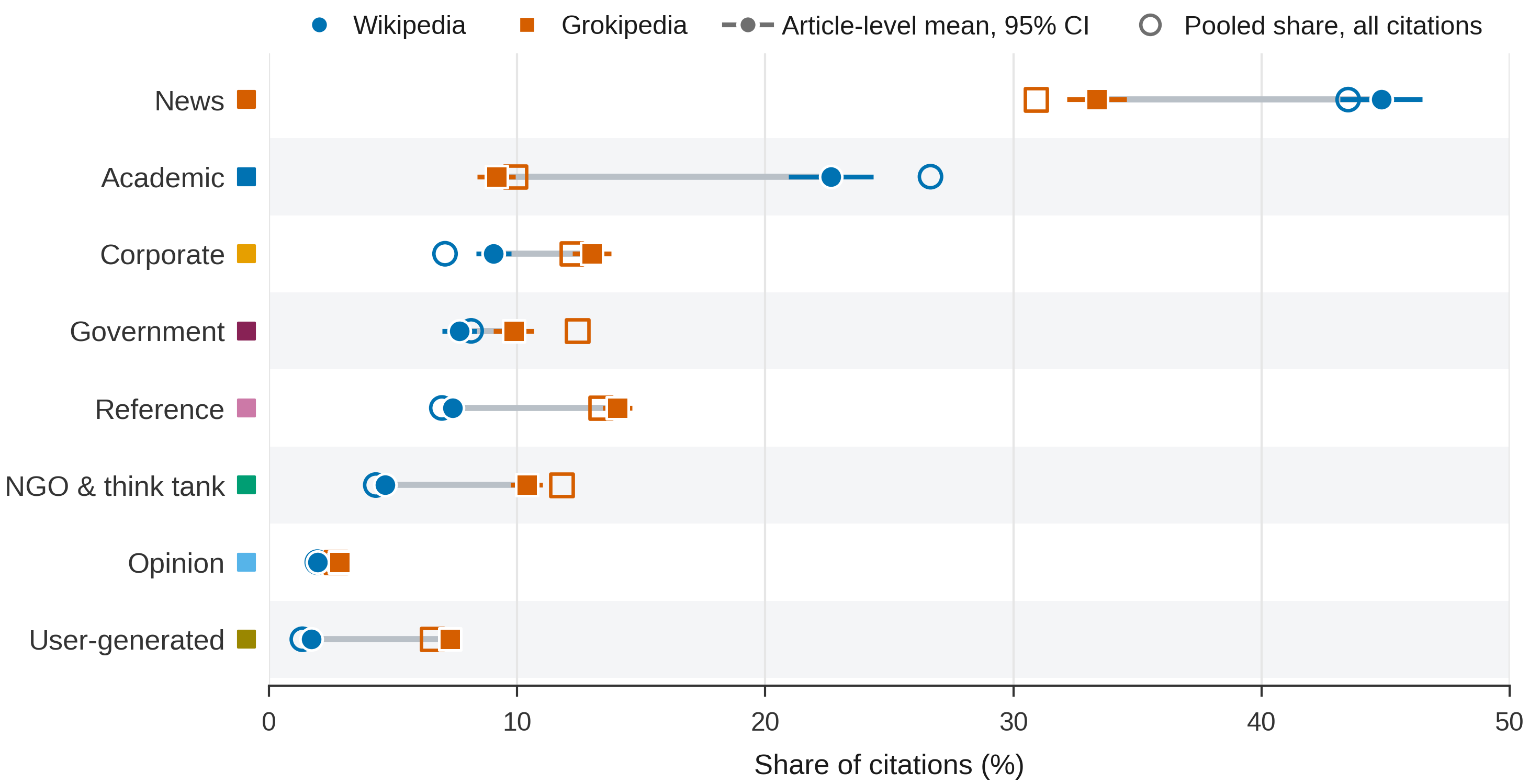}
  \caption{\textbf{Epistemic profiles of Wikipedia and Grokipedia.}
  Filled markers indicate the article-level mean share of each source category with 95\% confidence intervals ($N=1{,}000$ articles per platform). Open markers indicate the pooled share across all citations ($N_{\mathrm{Wiki}}=336{,}686$, $N_{\mathrm{Grok}}=257{,}486$).}

  \label{fig:fingerprint_summary_combined}
\end{figure}

Using the epistemic profile defined in Section~\ref{sec:intro} and operationalized in Section~\ref{sec:methods}, we compare the institutional composition of citations across Wikipedia and Grokipedia. We examine two complementary measures: the \textit{pooled composition}, calculated across all citations, and the \textit{article-level mean}, which gives equal weight to each article regardless of its citation count. Figure~\ref{fig:fingerprint_summary_combined} presents both measures. Raw citation counts are reported in Table~\ref{tbl:fingerprint_summary}.

The results reveal differences in the types of sources cited by the two encyclopedias. Wikipedia relies predominantly on \textit{News \& Journalism} and \textit{Academic \& Scholarly} sources, which together account for 70.1\% of its pooled citations (Table~\ref{tbl:fingerprint_summary}). These categories also have the highest average shares at the article level (Figure~\ref{fig:fingerprint_summary_combined}).

Grokipedia exhibits a different composition. Its share of \textit{News \& Journalism} sources is 30.9\%, compared with 43.5\% for Wikipedia, a difference of 12.6 percentage points. The contrast is more pronounced for \textit{Academic \& Scholarly} sources, whose share in Wikipedia is approximately 2.7 times that in Grokipedia. In contrast, Grokipedia has higher proportions of \textit{Corporate \& Commercial}, \textit{Reference \& Tertiary}, \textit{Government \& Official}, and \textit{Opinion \& Advocacy} sources, each representing 1.4--1.9 times its corresponding Wikipedia share. The largest relative differences occur in \textit{NGO \& Think Tank} sources (2.7 times) and \textit{User-Generated Content} (4.9 times).

Overall, these findings indicate that Wikipedia and Grokipedia draw on different distributions of institutional sources to support their claims, with Grokipedia relying proportionally less on academic and journalistic sources and more on several other source categories. In this comparison, the AI-generated encyclopedia differs from its human-curated counterpart not only in how its text is produced but also in the sources it cites.

\subsection{RQ3: Topic Dependence of Sourcing}

\begin{figure}[htbp]
  \centering
  \includegraphics[width=\textwidth]{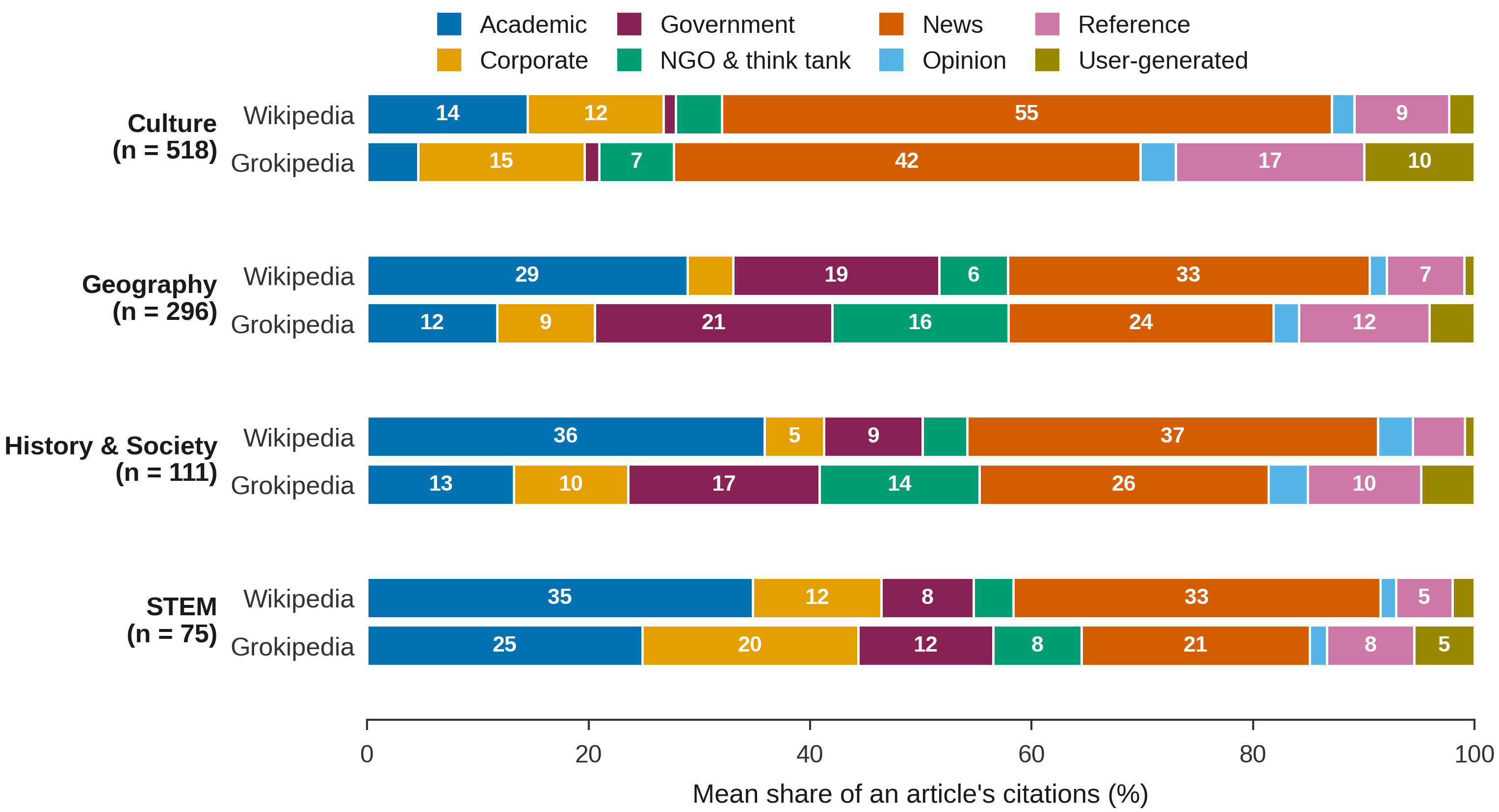}
  \caption[Epistemic profile by topic]{
  \textbf{Epistemic profiles across topic domains.}
  Mean article-level share of citations in each source category, displayed as 100\% stacked bars for Wikipedia and Grokipedia across four topical domains.
  Each domain shows Wikipedia above Grokipedia.
  Segments representing at least 6\% are labeled.
  Grokipedia has a lower share of \textit{Academic \& Scholarly} citations in all four domains.
  }
  \label{fig:topic_comparison}
\end{figure}

To examine how sourcing patterns vary by subject matter, we grouped the 1{,}000 matched article pairs into four broad domains using the Wikimedia topic taxonomy (Table~\ref{tab:corpus_topics}). Figure~\ref{fig:topic_comparison} compares the epistemic profiles of Wikipedia and Grokipedia across these domains, using the mean article-level share of each source category.

Both platforms exhibit topic-dependent sourcing patterns, but their epistemic profiles differ in the representation of academic sources. On Wikipedia, \textit{Culture} articles, including biographies, entertainment, and sports, rely predominantly on \textit{News \& Journalism} (55\% of citations in the average article), with a smaller share of \textit{Academic \& Scholarly} sources (14\%). In contrast, academic sources account for approximately one-third of citations in \textit{History \& Society} (36\%) and \textit{STEM} (35\%) articles.

Grokipedia's academic share is lower across all four domains: 14\% versus 36\% in \textit{History \& Society}, 12\% versus 29\% in \textit{Geography}, 25\% versus 35\% in \textit{STEM}, and less than 5\% versus 14\% in \textit{Culture}. The differences in other source categories vary by domain. In \textit{History \& Society}, Grokipedia has higher shares of \textit{Government \& Official} (17\% versus 9\%) and \textit{NGO \& Think Tank} sources (14\%). In \textit{Geography}, \textit{NGO \& Think Tank} sources account for 16\% of citations, compared with 6\% in Wikipedia. In \textit{Culture}, Grokipedia relies more on \textit{Reference \& Tertiary} sources (17\% versus 9\%) and \textit{User-Generated Content} (10\%). In \textit{STEM}, \textit{Corporate \& Commercial} sources represent 19\% of citations, compared with 11\% in Wikipedia.

These findings suggest that differences in the institutional composition of cited sources are not uniform across topics, although Grokipedia's lower reliance on academic sources is observed in every domain.

\subsubsection{Topic-Based Divergence and Evenness}

To examine whether differences in epistemic profiles vary across topics, we calculated the Jensen--Shannon distance between the citation distributions of each matched Wikipedia--Grokipedia article pair~\cite{cover2006elements}. This measure quantifies the dissimilarity between two probability distributions. Specifically, we use the square root of the Jensen--Shannon divergence, computed with natural logarithms. Higher values indicate greater differences in the composition of cited sources.

\begin{figure}[htbp]
  \centering
  \includegraphics[width=\textwidth]{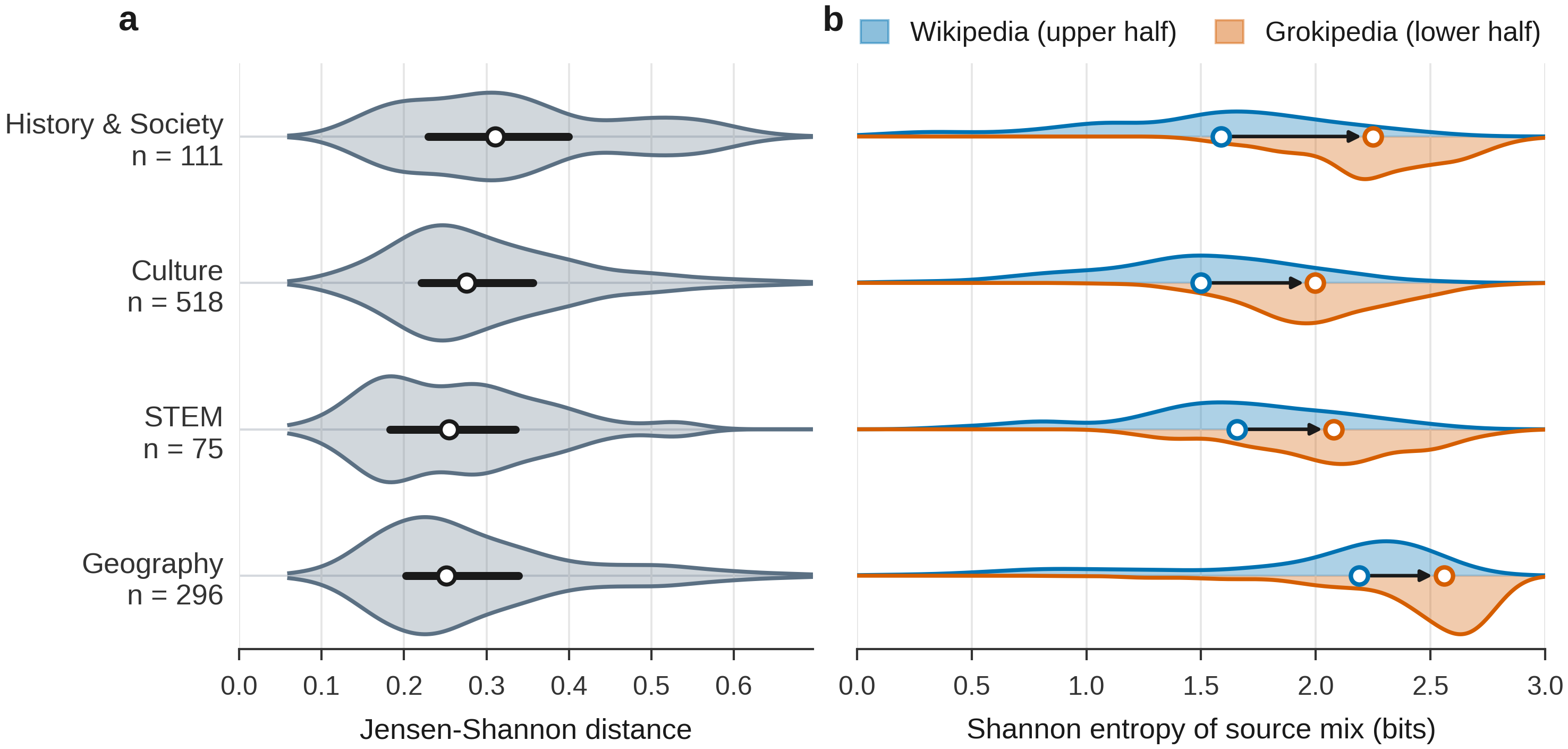}
  \caption[Topic-based epistemic metrics]{
  \textbf{Topic-based epistemic metrics.}
  \textbf{(a)} Jensen--Shannon distance between the citation-category distributions of matched Wikipedia and Grokipedia articles, grouped by topical domain. Density distributions are shown with interquartile ranges (bars) and medians (dots). Domains are ordered by median distance. Kruskal--Wallis $H=21.71$, $p<0.001$, $\eta^2_H=0.019$.
  \textbf{(b)} Shannon entropy of article-level citation-category distributions. Wikipedia distributions appear above each horizontal line and Grokipedia distributions below, with arrows connecting the medians. Paired effect sizes ($d_z$): Culture 1.16, Geography 1.00, History \& Society 1.25, and STEM 0.88.
  }
  \label{fig:topic_metrics}
  \label{fig:topic_metrics_jsd}
  \label{fig:topic_metrics_entropy}
\end{figure}

We compared these distances across the four topical domains using a Kruskal--Wallis H-test~\cite{kruskal1952use}, given the non-normal distribution of the scores and unequal group sizes. The results indicate statistically significant differences across domains ($H=21.71$, $p<0.001$, $\eta^2_H=0.019$~\cite{tomczak2014need}; Figure~\ref{fig:topic_metrics_jsd}). However, the effect is small, with topic accounting for approximately 2\% of the variation in ranks. \textit{History \& Society} exhibits the highest median distance (0.313), followed by \textit{Culture} (0.276), while \textit{Geography} and \textit{STEM} have the lowest (both 0.251). Despite these differences, the distributions overlap substantially, indicating considerable variation between individual article pairs within each domain.

We also examined the Shannon entropy of each article's citation-category distribution to assess the evenness of its epistemic profile~\cite{pielou1966measurement,magurran2013ecological}. Higher entropy indicates a more even distribution of citations across source categories, whereas lower entropy indicates greater concentration in a smaller number of categories. A more evenly distributed profile is not necessarily epistemically superior to one concentrated in academic or other specialized sources.

Grokipedia exhibits higher citation-category entropy than Wikipedia across all four domains (Figure~\ref{fig:topic_metrics_entropy}). To quantify these differences, we calculated paired Cohen's $d_z$ effect sizes. Because the paired entropy differences departed from normality (Shapiro--Wilk $p<0.001$), we also performed a Wilcoxon signed-rank test, which confirmed an overall difference ($p<0.001$; matched-pairs rank-biserial $r=0.94$; Appendix~\ref{sec:appendix_Effect Sizes}).

The effect sizes are large across all domains: $d_z=1.25$ [1.10, 1.44] for \textit{History \& Society}, 1.16 [1.06, 1.27] for \textit{Culture}, 1.00 [0.92, 1.09] for \textit{Geography}, and 0.88 [0.68, 1.13] for \textit{STEM} (Table~\ref{tab:entropy_stats}). Thus, although the magnitude varies by topic, Grokipedia exhibits a more even distribution of citations across source categories than Wikipedia.

\subsubsection{Article-by-Article Sourcing Divergence}

\begin{figure}[htbp]
  \centering
  \includegraphics[width=\textwidth]{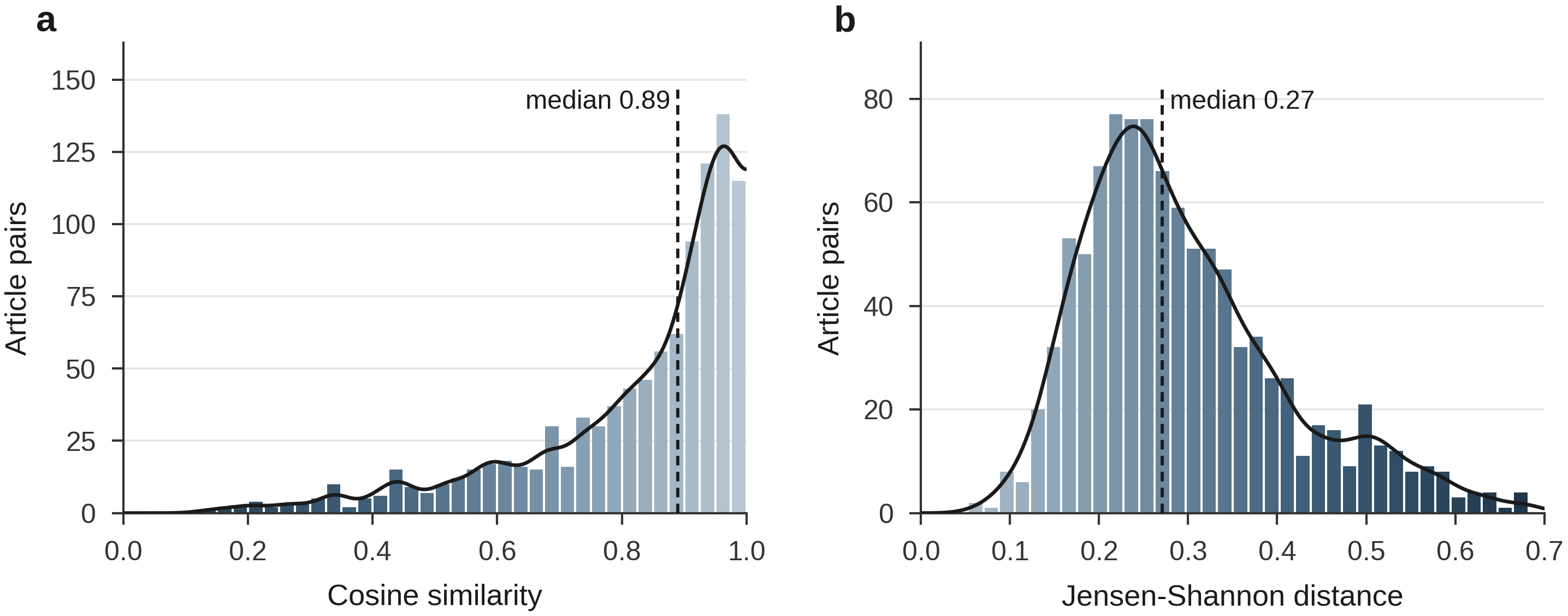}
  \caption[Per-article divergence between Wikipedia and Grokipedia]{
  \textbf{Article-level sourcing similarity between Wikipedia and Grokipedia}
  ($N=1{,}000$ matched article pairs).
  \textbf{(a)} Cosine similarity between the eight-category citation distributions of matched articles.
  \textbf{(b)} Jensen--Shannon distance between the corresponding epistemic profiles.
  Dashed lines indicate the medians.
  }
  \label{fig:divergence_histograms}
\end{figure}

To examine differences in sourcing at the article level, we compared the epistemic profiles of all 1{,}000 matched Wikipedia--Grokipedia article pairs. Each article was represented by an eight-dimensional probability vector containing the proportion of its citations assigned to each source category. We measured the similarity between paired profiles using two complementary metrics: Jensen--Shannon distance, which captures differences between probability distributions, and cosine similarity, which measures the alignment of their category proportions and which Mohammadi and Yasseri~\cite{mohammadi2026selective} also use to compare matched articles. The resulting distributions are shown in Figure~\ref{fig:divergence_histograms}.

The cosine similarity distribution is concentrated toward higher values, with a median of 0.89 and 47\% of article pairs exceeding 0.90. This indicates that many matched articles have similar proportions of source categories, despite the aggregate differences reported in RQ2. However, the distribution of Jensen--Shannon distances is right-skewed, revealing a smaller set of articles with substantially different sourcing profiles. These findings complement previous evidence of textual similarity between Wikipedia and Grokipedia~\cite{mohammadi2026selective}, while showing that similarities in content and citation composition represent distinct dimensions of comparison.

Table~\ref{tab:top_articles_combined_compact} presents the eight most similar and dissimilar pairs based on Jensen--Shannon distance. The most divergent examples include historical and political subjects, biographies, and a literary series, such as \textit{The Holocaust}, \textit{American Revolutionary War}, \textit{Ayn Rand}, and \textit{Warriors}. The most similar include geographic subjects, sports teams, and popular-culture articles, such as \textit{Switzerland}, \textit{San Francisco}, \textit{iPod}, and \textit{Paris Hilton}. These examples illustrate substantial variation at the article level, consistent with the earlier finding that broad topical domains explain only a small proportion of sourcing divergence.

\begin{table}[htbp]
\centering
\small
\caption[Most dissimilar and most similar article pairs]{
\textbf{Most dissimilar and most similar article pairs.}
The eight pairs with the highest and lowest Jensen--Shannon distances between Wikipedia's and Grokipedia's epistemic profiles, with corresponding cosine similarity (Cos.).
Topics follow the Wikimedia classification at the top and second levels.
}
\label{tab:top_articles_combined_compact}
\renewcommand{\arraystretch}{1.12}
\setlength{\tabcolsep}{6pt}
\begin{tabular}{@{} l l r r @{}}
\toprule
\textbf{Article} & \textbf{Topic} & \textbf{JS dist.} & \textbf{Cos.} \\
\midrule
\multicolumn{4}{@{}l}{\textit{(a) Most dissimilar}} \\
\addlinespace[2pt]
Warriors (novel series) & Culture $\cdot$ Literature & 0.68 & 0.13 \\
American Revolutionary War & Geography $\cdot$ Regions & 0.68 & 0.18 \\
The Holocaust & Geography $\cdot$ Regions & 0.68 & 0.20 \\
Ayn Rand & Culture $\cdot$ Philosophy and religion & 0.68 & 0.16 \\
Marilyn Monroe & Culture $\cdot$ Media & 0.65 & 0.16 \\
George Washington & Geography $\cdot$ Regions & 0.64 & 0.21 \\
Joan of Arc & Culture $\cdot$ Biography & 0.64 & 0.42 \\
Vladimir Lenin & History \& Society $\cdot$ Politics and government & 0.64 & 0.47 \\
\midrule
\multicolumn{4}{@{}l}{\textit{(b) Most similar}} \\
\addlinespace[2pt]
Switzerland & Geography $\cdot$ Regions & 0.07 & 0.99 \\
Tropical cyclone & STEM $\cdot$ STEM & 0.07 & 1.00 \\
San Francisco & Geography $\cdot$ Regions & 0.09 & 0.99 \\
Philadelphia & Geography $\cdot$ Regions & 0.09 & 0.98 \\
iPod & Culture $\cdot$ Media & 0.09 & 0.99 \\
Greece national football team & Culture $\cdot$ Sports & 0.10 & 0.98 \\
Paris Hilton & Culture $\cdot$ Media & 0.10 & 1.00 \\
Colombia national football team & Culture $\cdot$ Sports & 0.10 & 1.00 \\
\bottomrule
\end{tabular}
\end{table}

\subsection{The Core Epistemic Network}

\begin{figure}[htbp]
  \centering
  \begin{minipage}[b]{0.48\textwidth}
    \centering
    \includegraphics[width=\textwidth]{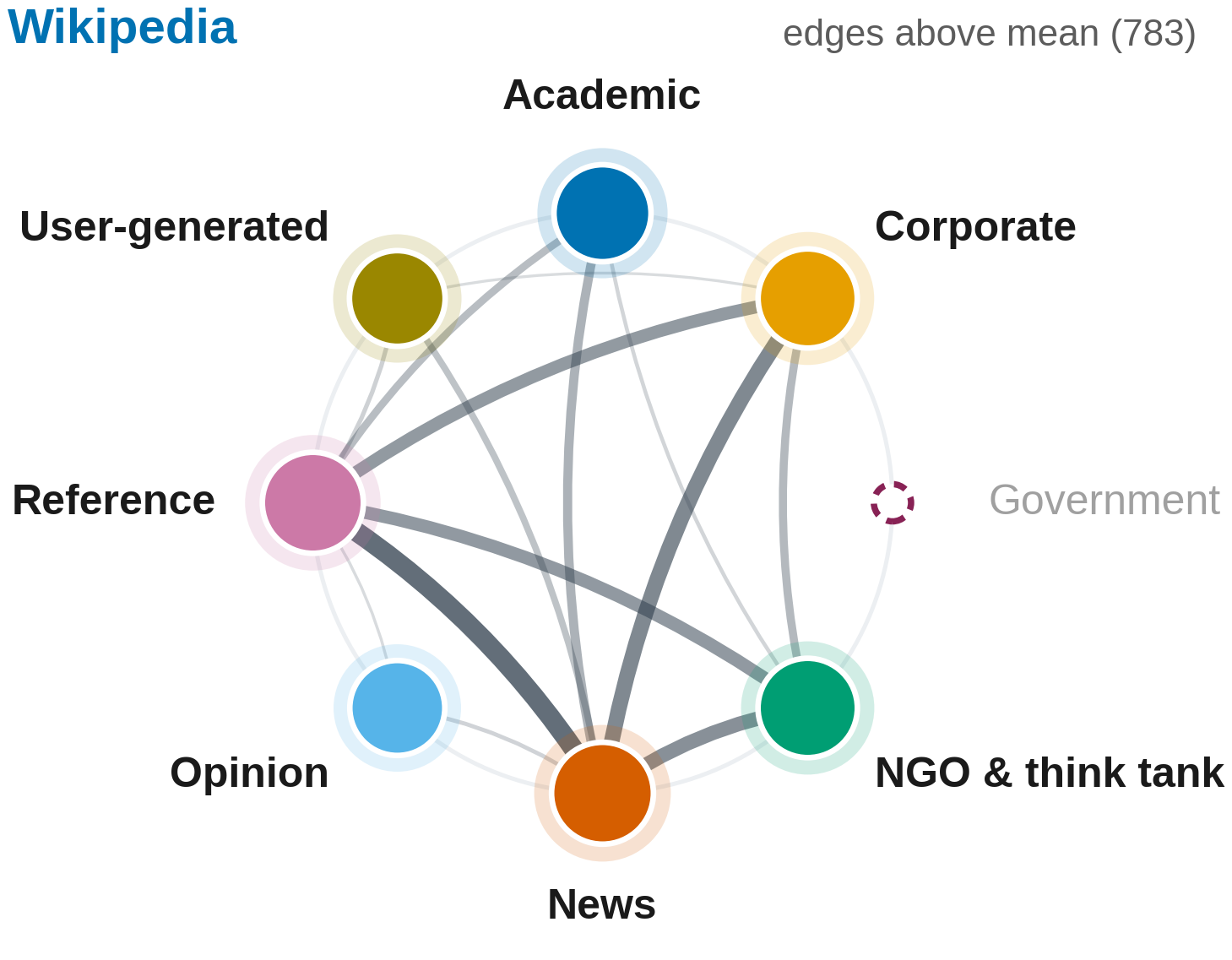}
    \caption*{(a) Wikipedia (Core)}
  \end{minipage}
  \hfill
  \begin{minipage}[b]{0.48\textwidth}
    \centering
    \includegraphics[width=\textwidth]{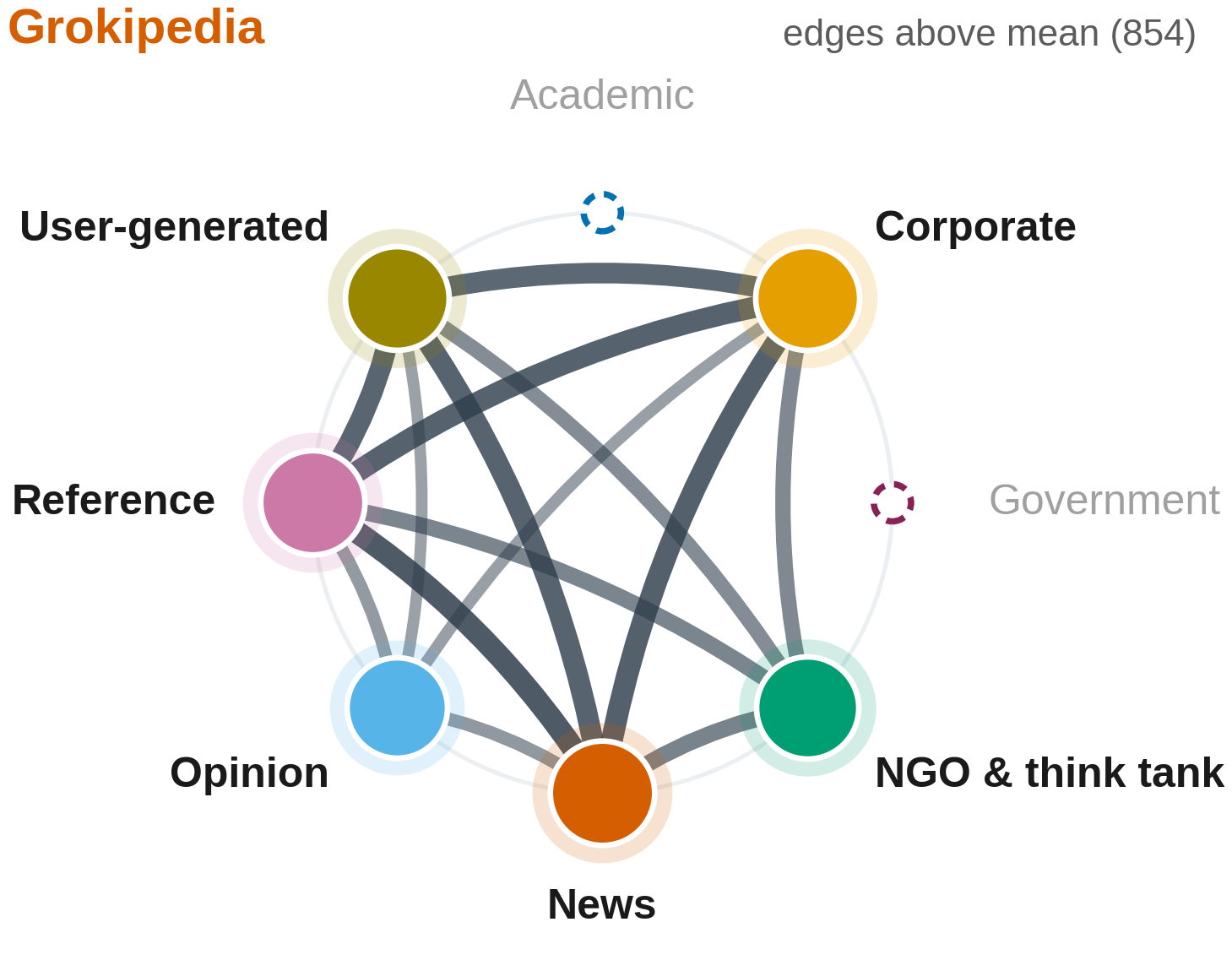}
    \caption*{(b) Grokipedia (Core)}
  \end{minipage}

  \caption{\textbf{Core epistemic co-occurrence networks.}
  Nodes represent source categories, and edges indicate their co-occurrence within articles. Only edges with weights above the platform-specific mean are retained (Wikipedia: 782.8; Grokipedia: 854.2). Darker and thicker edges indicate stronger co-occurrence.
  \textbf{(a)} Wikipedia.
  \textbf{(b)} Grokipedia.}
  \label{fig:co_occurrence_nets}
\end{figure}

To visualize how different source categories co-occur within articles, we constructed filtered co-occurrence networks for Wikipedia and Grokipedia (Figure~\ref{fig:co_occurrence_nets}). Each node represents one of the eight epistemic categories, and each edge is weighted by the number of articles in which the two categories appear together. A category was considered present if an article contained at least one citation from that category. To highlight the strongest connections, we retained only edges with weights strictly above the mean co-occurrence weight calculated separately for each platform.

These networks provide a descriptive view of the relationships between source categories rather than independent statistical evidence. The mean-weight threshold is a relative filtering criterion. To assess sensitivity to the filtering choice, we repeated the analysis using the median, 60th percentile, and 75th percentile of edge weights. We also examined a cosine-normalized network built from the presence or absence of each category in each article (Appendix~\ref{sec:appendix_sensitivity}).

The networks reveal different patterns of source-category co-occurrence. Wikipedia exhibits an interconnected core, with \textit{Academic \& Scholarly} sources linked to \textit{News \& Journalism}, \textit{Reference \& Tertiary}, and \textit{NGO \& Think Tank}. Its strongest connection is between \textit{News \& Journalism} and \textit{Reference \& Tertiary} sources.

In Grokipedia, by contrast, \textit{Academic \& Scholarly} sources have no connections above the mean threshold and are therefore absent from the connected core. This finding is robust across the alternative thresholds examined. Meanwhile, \textit{User-Generated Content} appears in both networks but participates in some of Grokipedia's strongest connections, particularly with \textit{Corporate \& Commercial} and \textit{News \& Journalism} sources. \textit{Government \& Official} sources fall below the threshold on both platforms.

\subsection{Article Homophily Network}

\begin{figure}[htbp]
  \centering
  \includegraphics[width=\textwidth]{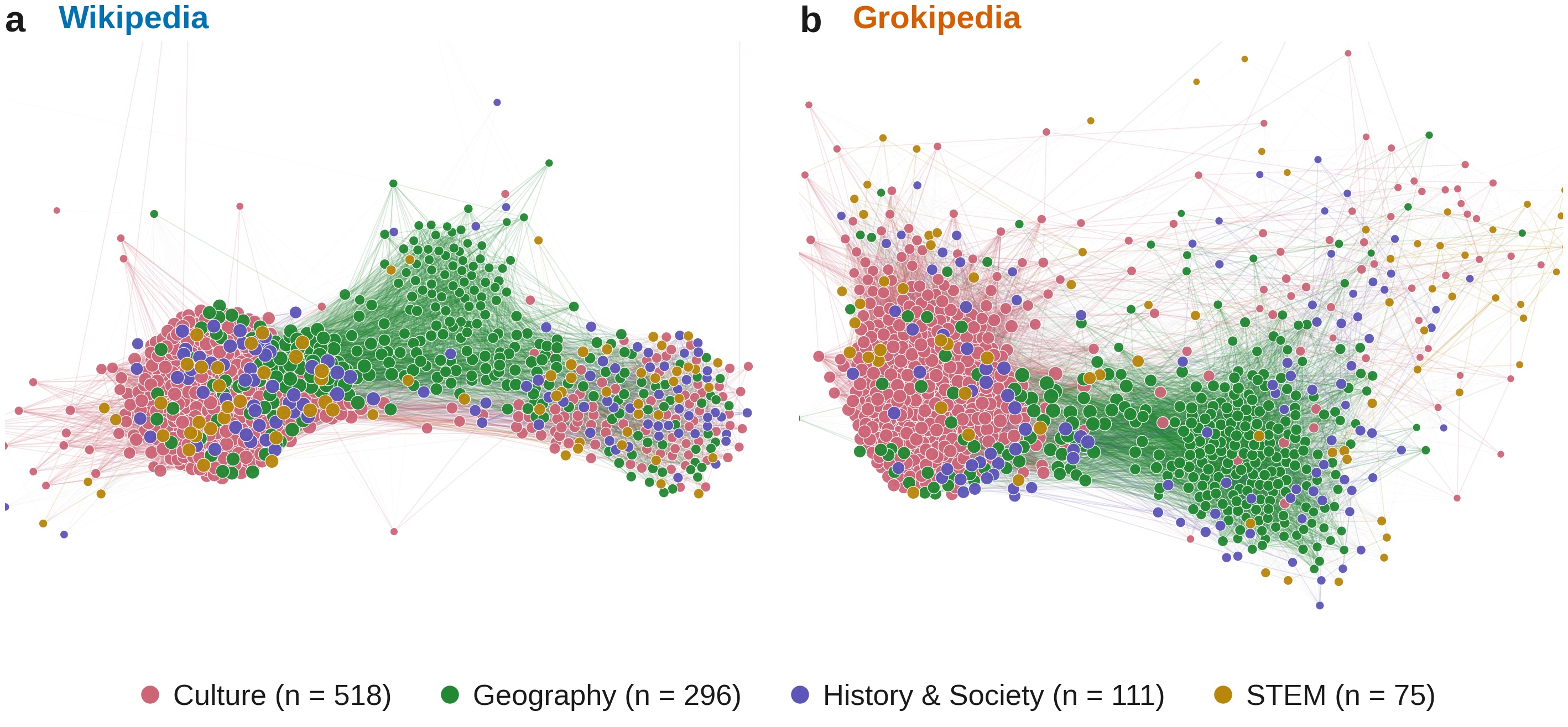}
  \caption[Article homophily networks]{
  \textbf{Article similarity networks (cosine similarity $>0.75$).}
  Each node represents an article, colored by topic domain and sized by degree. Edges connect articles whose epistemic profiles have cosine similarity above 0.75. Within-topic edges take the corresponding topic color, while between-topic edges are gray. Each network is visualized using a force-directed layout, at most 25{,}000 randomly sampled edges are drawn.
  \textbf{(a)} Wikipedia.
  \textbf{(b)} Grokipedia.
  }
  \label{fig:similarity_nets}
\end{figure}

To examine whether articles with similar epistemic profiles also share topical characteristics, we constructed article similarity networks for Wikipedia and Grokipedia. Each of the 1{,}000 articles per platform was represented by an eight-dimensional epistemic profile, and cosine similarity was calculated between every pair of articles within each platform. We connected two articles when their similarity exceeded 0.75, producing the networks shown in Figure~\ref{fig:similarity_nets}. This threshold was chosen because it exceeds the mean pairwise similarity of 0.64 on both platforms while preserving a single connected component in each network.

Dichotomizing continuous similarity scores can introduce arbitrary structural features~\cite{maccallum2002practice,altman2006cost}. We therefore use the thresholded networks primarily for visualization and assess the robustness of the findings across thresholds from 0.60 to 0.95 (Appendix~\ref{sec:appendix_e}; Table~\ref{tab:assortativity_sensitivity}). We also conduct a threshold-free analysis using the original similarity matrices.

The visualizations suggest differences in how articles with similar sourcing profiles are distributed across topics. Grokipedia shows a more pronounced separation between regions dominated by \textit{Culture} and \textit{Geography} articles, while Wikipedia exhibits greater mixing across domains. Because force-directed layouts are sensitive to visualization choices, we evaluate these patterns using network statistics rather than relying on visual inspection alone.

Topic assortativity provides the clearest distinction. At the 0.75 threshold, Grokipedia exhibits higher assortativity ($r=0.297$) than Wikipedia ($r=0.153$), indicating a stronger tendency for articles with similar sourcing profiles to belong to the same topical domain. This pattern persists across thresholds from 0.60 to 0.95, where Grokipedia's assortativity is approximately 1.7--2.4 times that of Wikipedia (Appendix~\ref{sec:appendix_e}; Table~\ref{tab:assortativity_sensitivity}).

Other network measures provide a complementary view (Table~\ref{tab:network_stats}). At the 0.75 threshold, Grokipedia's network is sparser than Wikipedia's (density 0.39 versus 0.46) and has lower transitivity (0.81 versus 0.87), but slightly higher modularity (Louvain $Q=0.310$ versus $0.291$). These differences are broadly consistent across the examined thresholds.

\begin{table}[htbp]
\centering
\small
\caption[Network statistics of the article homophily networks]{
\textbf{Article similarity network statistics across cosine similarity thresholds} ($N=1{,}000$ articles per platform).
Density is the proportion of possible edges present; Comp. denotes connected components; Giant is the proportion of articles in the largest component; Trans. is transitivity; $Q$ is Louvain modularity; Comm. is the number of Louvain communities containing more than one article; and $r$ denotes topic assortativity. All assortativity coefficients are positive and statistically significant under 1{,}000 topic-label permutations ($p<0.001$).
}
\label{tab:network_stats}
\renewcommand{\arraystretch}{1.12}
\setlength{\tabcolsep}{4.5pt}
\begin{tabular}{@{} l r r r r r r r r @{}}
\toprule
 & \multicolumn{2}{c}{\textbf{Size}}
 & \multicolumn{5}{c}{\textbf{Structure}}
 & \multicolumn{1}{c}{\textbf{Topic}} \\
\cmidrule(lr){2-3}
\cmidrule(lr){4-8}
\cmidrule(l){9-9}
\textbf{Platform} & Edges & Density & Comp. & Giant & Trans. & $Q$ & Comm. & $r$ \\
\midrule
\multicolumn{9}{@{}l}{\textit{Threshold 0.65}} \\
\quad Wikipedia
 & 278{,}021 & 0.56 & 1 & 1.000 & 0.87 & 0.246 & 2 & 0.105 \\
\quad Grokipedia
 & 261{,}025 & 0.52 & 1 & 1.000 & 0.81 & 0.247 & 2 & 0.195 \\
\addlinespace
\multicolumn{9}{@{}l}{\textit{Threshold 0.75}} \\
\quad Wikipedia
 & 230{,}423 & 0.46 & 1 & 1.000 & 0.87 & 0.291 & 3 & 0.153 \\
\quad Grokipedia
 & 195{,}128 & 0.39 & 1 & 1.000 & 0.81 & 0.310 & 2 & 0.297 \\
\addlinespace
\multicolumn{9}{@{}l}{\textit{Threshold 0.85}} \\
\quad Wikipedia
 & 170{,}078 & 0.34 & 1 & 1.000 & 0.87 & 0.354 & 3 & 0.200 \\
\quad Grokipedia
 & 124{,}032 & 0.25 & 2 & 0.999 & 0.79 & 0.376 & 4 & 0.405 \\
\bottomrule
\end{tabular}
\end{table}

To assess statistical significance, we randomly permuted article topic labels 1,000 times while preserving network connections and topic frequencies. Topic assortativity was significantly higher than expected under random labeling for both platforms ($p<0.001$). To verify that this pattern was not driven by the chosen similarity threshold, we also analyzed the full, unthresholded cosine similarity matrices. The mean similarity between articles within the same topic exceeded that between articles from different topics by 0.176 in Grokipedia and 0.126 in Wikipedia (both $p<0.001$, based on 1,000 label permutations). These results indicate a stronger observed association between topic membership and sourcing-profile similarity in Grokipedia, independent of the network threshold. A closer examination of topic mixing patterns shows that Grokipedia's higher assortativity is primarily associated with stronger within-topic connectivity in \textit{Culture} and \textit{Geography}, rather than a uniform separation across all domains (Appendix~\ref{sec:appendix_mixing}).

\subsection{Grokipedia over Time}
\label{sec:grok_time}

We close with a comparison of two Grokipedia snapshots taken ten months apart. The preceding analyses rely on a single snapshot, leaving open whether Grokipedia's sourcing patterns change over time. To examine this, we compared 69 articles captured at two points approximately ten months apart: November 2025, shortly after Grokipedia's launch, and September 2026.

\begin{figure}[htbp]
  \centering
  \includegraphics[width=\textwidth]{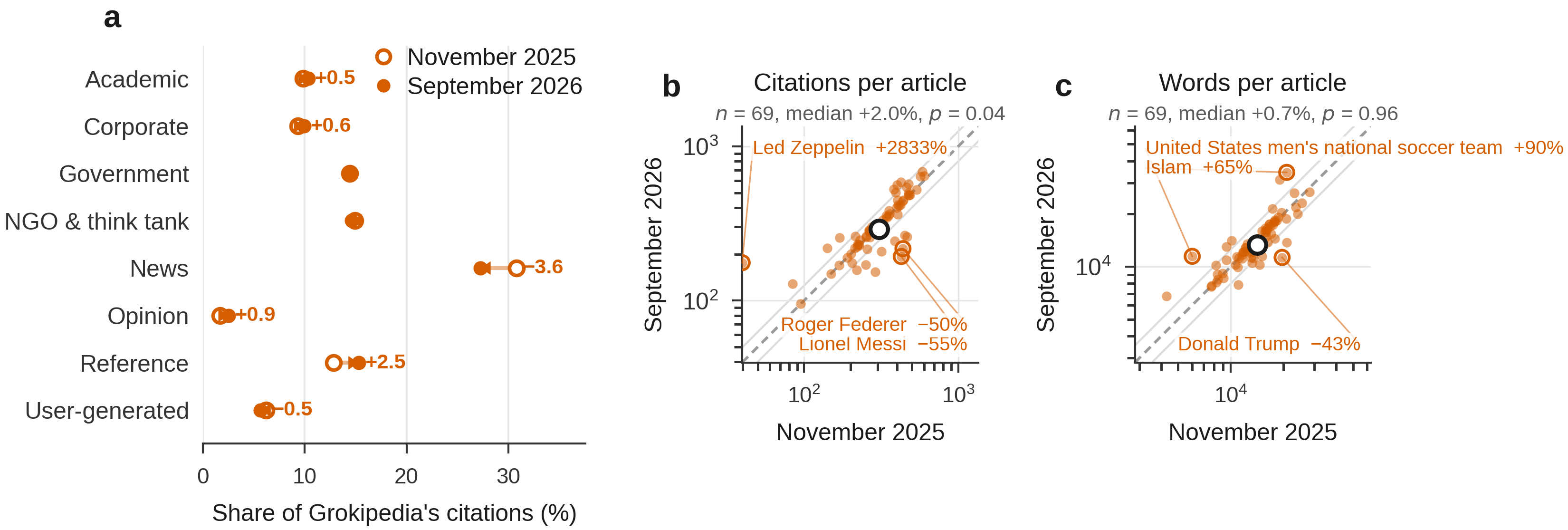}
  \caption[Grokipedia between November 2025 and September 2026]{
  \textbf{Changes in Grokipedia between November 2025 and September 2026}
  for 69 articles observed at both dates.
  \textbf{(a)} Pooled citation composition across the eight source categories, with percentage-point changes between observations.
  \textbf{(b)} Citation counts and
  \textbf{(c)} article lengths at the two dates.
  Dashed lines indicate equality, faint lines indicate changes of $+25\%$ and $-20\%$, the open black circles represent the marginal medians, and rings highlight the articles with the largest changes in each measure.
  }
  \label{fig:grok_time}
\end{figure}

Overall, article length and citation density changed little, although some individual articles underwent substantial revisions. The median citation count declined from 306 to 290, while the median within-article percentage change was $+2.0\%$ ($p=0.038$). Median changes in article length ($+0.7\%$, $p=0.96$) and citation density ($+0.1\%$, $p=0.58$) were not statistically significant (Table~\ref{tab:grok_time}). Notable exceptions include \textit{Lionel Messi}, whose citations declined from 427 to 193, and \textit{Roger Federer}, from 438 to 217 (Figure~\ref{fig:grok_time}b). Meanwhile, the \textit{Islam} article increased in length from 20{,}688 to 34{,}458 words, and the \textit{United States men's national soccer team} article grew by approximately 90\% (Figure~\ref{fig:grok_time}c).

Changes in citation composition were more apparent. The median Jensen--Shannon distance between each article's November and September epistemic profiles was 0.101. In the pooled citation distribution, \textit{News \& Journalism} declined by 3.6 percentage points, while \textit{Reference \& Tertiary} increased by 2.5 points. Smaller increases occurred in \textit{Opinion \& Advocacy} ($+0.9$), \textit{Corporate \& Commercial} ($+0.6$), and \textit{Academic \& Scholarly} ($+0.5$, from 9.9\% to 10.4\% within the 69-article panel) (Figure~\ref{fig:grok_time}a). Thus, even as overall citation density remained relatively stable, the composition of cited sources changed.

We also observed citations to Grokipedia itself. While none appeared in the November reference lists of the 69 longitudinally sampled articles, the larger September 2026 corpus contained 269 citations to \texttt{grokipedia.com} across 29 of the 1{,}000 articles, including topics such as \textit{Iran} and \textit{Israel}. Although these observations come from different sample sizes, they identify self-citation as an additional feature of Grokipedia's sourcing practices.

\begin{table}[htbp]
\centering
\small
\caption[Grokipedia between the two dates]{
\textbf{Longitudinal changes in Grokipedia articles.}
Comparison of 69 matched articles observed in November 2025 and September 2026. Median percentage changes are calculated at the article level. Reported $p$-values are from paired Wilcoxon signed-rank tests. Jensen--Shannon distance measures changes in each article's citation-category distribution between the two dates and is reported as median [interquartile range].
}
\label{tab:grok_time}
\renewcommand{\arraystretch}{1.15}
\setlength{\tabcolsep}{7pt}
\input{tables/tab_grok_time}
\end{table}


\section{Discussion}

Our findings point to a structural difference in the sources that Wikipedia and Grokipedia invoke to justify encyclopedic claims. We distinguish between what our analysis directly demonstrates and its broader epistemological implications. Empirically, we document differences in citation density, institutional composition, and topic-dependent sourcing patterns. We interpret these differences as evidence of \textit{epistemic substitution}: a change in the composition of cited institutional authority between articles covering the same topics. This interpretation rests on the premise that citations provide observable traces of the external authority invoked to substantiate knowledge claims. It does not establish how readers evaluate these sources or whether the resulting claims are accurate.

Beginning with sourcing quantity (RQ1), Grokipedia exhibits an approximately linear relationship between article length and citation count, adding roughly 21 citations per additional 1{,}000 words. Wikipedia, by contrast, exhibits greater variation in citation density and a nonlinear relationship in which citation growth slows for longer articles. Our results suggest that the two platforms follow different patterns of citation allocation. Grokipedia's relatively uniform scaling is consistent with a mechanical quota, in which citations are added at a roughly fixed rate per unit of text regardless of subject, whereas Wikipedia's greater variation is consistent with heterogeneous editorial practices across articles. However, our analysis cannot establish whether these differences arise from deliberate editorial rules, automated generation procedures, or other factors.

Turning to institutional composition (RQ2), the most notable difference is Grokipedia's lower reliance on \textit{Academic \& Scholarly} sources. Wikipedia draws more heavily on academic and journalistic publications, whereas Grokipedia cites proportionally more \textit{NGO \& Think Tank}, \textit{User-Generated Content}, and several other source categories. These differences matter because different source types are subject to different forms of institutional oversight. Scholarly publications, professional journalism, government documents, and user-generated content follow different processes of production and verification~\cite{wilson1983second,rieh2002judgment}. However, these categories do not constitute a hierarchy of reliability, nor does academic provenance guarantee factual accuracy. What our findings establish is a difference in the institutional composition of the sources invoked to support encyclopedic claims, rather than a difference in their demonstrated truthfulness.

These differences are also topic-dependent (RQ3). Grokipedia's academic citation share is lower across all four topical domains, while its source-category entropy is consistently higher, indicating a more even distribution across source types. The largest differences in epistemic profiles occur in \textit{History \& Society}, whereas \textit{STEM} exhibits comparatively smaller differences. The network analysis provides a complementary perspective: articles with similar sourcing profiles are more strongly associated with the same topic in Grokipedia than in Wikipedia, as indicated by higher topic assortativity and a larger within-topic similarity advantage. However, topic explains only a small proportion of the variation in article-level divergence ($\eta^2_H=0.019$). Thus, sourcing differences are observable across domains, but much of their variation occurs between individual articles rather than between broad topic categories.

The longitudinal comparison further suggests that Grokipedia's sourcing practices are evolving. Across 69 articles observed in November 2025 and September 2026, overall citation density remained relatively stable, while the distribution of cited source categories changed. We also identified references to Grokipedia itself in the larger September corpus. Although the longitudinal sample is limited, these observations reinforce the importance of treating an encyclopedia's epistemic profile as potentially dynamic rather than fixed at launch.

A further question concerns why these differences emerge. Grokipedia's internal retrieval procedures, generation instructions, and source-selection mechanisms are not publicly documented in sufficient detail for us to attribute these patterns to any particular mechanism. The observed differences may reflect model behavior, retrieval configurations, editorial instructions, or combinations of these factors. Distinguishing among these explanations requires additional evidence.

Taken together, our results suggest that the transition from human-curated to AI-generated encyclopedic content involves more than a change in how text is produced. It is accompanied, in this comparison, by differences in the institutional sources invoked to support knowledge claims. Whether such differences affect information quality, public trust, or users' understanding remains an open empirical question. Nevertheless, the ability to observe and compare these structures provides a basis for auditing the epistemic foundations of generative knowledge systems.

\section{Conclusion}

This study examined how human-curated and AI-generated encyclopedias differ in the sources they cite to justify knowledge claims. Using 1{,}000 matched Wikipedia--Grokipedia article pairs, we developed an eight-category epistemic classification scheme and compared citation density, institutional composition, and topic-dependent sourcing patterns.

Our findings address three research questions. First (RQ1), Grokipedia exhibits a more uniform relationship between article length and citation count, while Wikipedia's citation density varies more substantially across articles. Second (RQ2), the two platforms differ in their institutional composition of sources, with Grokipedia relying proportionally less on academic and journalistic publications and more on several other source categories. Third (RQ3), these differences vary across topics, with Grokipedia exhibiting greater source-category evenness and stronger topic assortativity in its article similarity networks.

Beyond these empirical findings, our contribution is methodological. We introduce the \textit{Epistemic Profile} as a way to characterize the distribution of institutional authority invoked through citations. Drawing on Buckland's concept of \textit{information-as-thing}, we treat citations as observable informational artifacts whose provenance can be systematically classified and compared~\cite{buckland1991information}. This approach complements existing evaluations of factual accuracy, citation reliability, and political bias by examining a different dimension: the institutional composition of the sources used to substantiate claims.

The framework also provides a basis for auditing other AI-mediated information systems, including search engines, recommendation systems, and retrieval-augmented generation pipelines. Such applications would require adaptation and validation for their respective source environments. More broadly, external audits can reveal systematic patterns in algorithmic outputs even when the underlying computational processes remain inaccessible~\cite{lazer2015rise,rahwan2019machine,sandvig2014auditing,raji2020closing,wagner2021measuring,tutt2017fda,hilbert20248}.

Information science, with its longstanding concern for how knowledge is produced, organized, evaluated, and made accessible~\cite{hjorland2002domain,bates1999invisible}, is well positioned to investigate these emerging forms of information production. Our findings concern one AI-generated encyclopedia and should not be generalized to all generative systems. Nevertheless, they demonstrate that differences in the institutional foundations of cited knowledge can be systematically measured, making provenance-level auditing an important component of accountability for AI-generated information.

\section{Limitations and Future Work}

Several limitations should be considered when interpreting these findings. First, our corpus consists of 1{,}000 matched articles selected from highly revised English Wikipedia topics. Although this sampling strategy provides a substantial collection of prominent and frequently edited articles, it is not representative of either encyclopedia as a whole. Highly revised articles may differ from typical articles in editorial attention, sourcing practices, and topic composition. Some may also be subject to editing restrictions. Consequently, the magnitude and direction of the observed differences should be interpreted within this sampling frame. A replication on Wikipedia's Vital Articles, a corpus selected for importance rather than editing activity, shows the same direction of differences (Appendix~\ref{sec:appendix_vital}). Because highly revised articles may be sourced more strictly on Wikipedia than typical articles, the differences we observe could be larger or smaller in a representative sample, and our design does not indicate which. Future research should examine broader and more representative article samples, including additional languages, less frequently revised topics, and curated sets such as Wikipedia's Featured Articles, which are citation-rich but selected for quality rather than editing activity.

Second, Grokipedia is an evolving platform built on proprietary generative AI systems. Its retrieval, generation, and citation-selection procedures may change over time, limiting the temporal generalizability of our findings. Although our longitudinal comparison provides initial evidence of changes in sourcing composition, the sample is too limited to characterize the evolution of the platform as a whole. Repeated audits are therefore needed to assess whether the observed differences persist, diminish, or become more pronounced as the system develops~\cite{tutt2017fda}.

Third, citation classification relies partly on automated LLM annotation. While the primary and secondary classifiers agreed on 81.8\% of the validation sample, disagreements between models frequently coincided with difficulties in human classification. In particular, human agreement was substantially lower for citations on which the models disagreed. These results highlight ambiguity at the boundaries between certain source categories, such as \textit{Opinion \& Advocacy} and \textit{News \& Journalism}. Although our independent human validation provides evidence about classification consistency, it does not establish error-free classification or a definitive accuracy estimate for the entire corpus.

Fourth, our analysis examines citation structure rather than citation correctness or claim-level support. A citation may point to an authentic source without adequately supporting the accompanying claim, and similar source distributions do not necessarily imply similar factual content. We also did not systematically verify the existence and authenticity of every cited work. Previous research has documented fabricated or otherwise invalid citations in LLM-generated text~\cite{agrawal2024language,zhao2026llm}. URL accessibility alone cannot establish source authenticity, since a functioning link may lead to an incorrect source and an inaccessible link may refer to a legitimate publication. Future work could combine epistemic profiling with citation-grounding evaluations, and systematic source verification.

Fifth, Wikipedia should not be treated as an unbiased or definitive epistemic standard. Its content and sourcing practices are shaped by inequalities in editorial participation, including gender gaps among editors~\cite{hill2013wikipedia} and in how women are portrayed~\cite{wagner2015s}, geographic and linguistic imbalances~\cite{graham2014uneven}, conflicts over contested articles~\cite{yasseri2012dynamics}, and undisclosed paid editing~\cite{joshi2020detecting}. Its verifiability requirements also privilege previously published sources, potentially reproducing existing inequalities in access to institutional publication. Our findings therefore characterize differences in cited institutional authority, not the superiority of one encyclopedia's epistemic practices.

Finally, our observational design cannot identify the mechanisms responsible for the observed sourcing differences. Although matching articles by topic controls for differences in subject matter, other factors, including editorial histories, source availability, article length, retrieval practices, and human interventions, may influence citation patterns. Moreover, Wikipedia and Grokipedia should not necessarily be treated as independent knowledge systems. Grokipedia draws on Wikipedia content and incorporates human suggestions through Grok-mediated review, while recent research suggests that Grokipedia's introduction has also influenced editing activity on Wikipedia~\cite{leung2026ai}. These interactions point to a potentially interdependent knowledge ecosystem in which human and AI-generated content may influence one another. However, our analysis does not directly examine these feedback mechanisms or their effects on sourcing practices. Nor should differences in epistemic profiles be interpreted as evidence of changes in public trust, knowledge acquisition, or persuasion.

Several questions raised by this study are best addressed by merging our design with existing work. First, combining provenance profiles with the text-similarity measures of Mohammadi and Yasseri~\cite{mohammadi2026selective}, Triedman and Mantzarlis~\cite{triedman2025did}, and Eibl et al.~\cite{eibl2026grokipedia} would show, pair by pair, whether articles whose wording diverges are also those whose sources diverge. Second, Hadad et al.~\cite{hadad2026wikipedia} model Grokipedia's selection and rewriting of Wikipedia articles as a function of page views, editing activity, reverts, and reference density; applying the same covariates, together with established measures of controversy on Wikipedia~\cite{yasseri2012dynamics}, to provenance-level divergence would test whether the differences we observe reflect the salience or contestedness of articles. Third, Grokipedia lets readers submit edit requests that Grok reviews, and Hadad et al. have begun to analyse these requests; linking accepted and rejected requests to changes in citation profiles would show whether human intervention alters how articles are sourced, although the platform's review process has changed repeatedly since launch~\cite{robinson2026grok,diresta2026grokipedia}. Fourth, Wikipedia's policy on biographies of living persons~\cite{wikipedia_BLP} may shape how biographies are sourced, but so does the literature available about a person, which changes after death~\cite{keegan2015is,west2021postmortem}; comparing living and deceased subjects on both platforms, for example within the panel of biographies assembled by Leung and Strumpf~\cite{leung2026ai}, would separate the two.

\textbf{Supplementary Information:} Supplementary material relevant to this study is provided in the appendix.

\textbf{Data Sharing and Data Availability}

The data and code supporting the findings of this study will be made openly available upon publication of this article at \url{https://doi.org/10.5281/zenodo.23425913} and \url{https://github.com/faezehkhazaee97/epistemic-substitution}, respectively.

\textbf{Acknowledgments:} The uses and roles of generative AI in this study are declared in the Methods section.

\textbf{Conflict of Interest:} The authors declare no known competing financial interests or personal relationships that could have influenced the work reported in this paper.

\textbf{Author Contributions:} AM conceptualized the study and wrote the manuscript; AM and FK collected the data and conducted the analysis; MH supervised the project, developed the LLM content classifiers, and reviewed and edited the manuscript.

\bibliographystyle{unsrt}   
\bibliography{references}  

\begin{appendices}

\newpage

\section*{Appendix}
\setcounter{secnumdepth}{2}
\setcounter{section}{0}
\renewcommand{\thesection}{\Alph{section}}
\renewcommand{\thesubsection}{\thesection.\arabic{subsection}}

\setcounter{table}{0}
\setcounter{figure}{0}

\renewcommand{\thetable}{S\arabic{table}}
\renewcommand{\thefigure}{S\arabic{figure}}

\section{Composition of the Citation Base}
\label{sec:appendix_pooled}

\begin{table}[htbp]
\centering
\small
\caption[Global epistemic profile]{\textbf{Global epistemic profile, pooled across all citations.} Citation counts and corpus-level shares for Wikipedia and Grokipedia. Values give the total composition of the dataset ($N_{\mathrm{Wiki}}=336{,}686$, $N_{\mathrm{Grok}}=257{,}486$) rather than the per-article average.}
\label{tbl:fingerprint_summary}
\renewcommand{\arraystretch}{1.18}
\setlength{\tabcolsep}{8pt}
\begin{tabular}{@{} c l r r r r @{}}
\toprule
 & & \multicolumn{2}{c}{\textbf{Wikipedia}} & \multicolumn{2}{c}{\textbf{Grokipedia}} \\
\cmidrule(lr){3-4}\cmidrule(l){5-6}
 & \textbf{Category} & $n$ & \% & $n$ & \% \\
\midrule
{\color{cAcad}\swbox} & Academic \& Scholarly
  & 89{,}763 & 26.66
  & 25{,}574 & 9.93 \\
{\color{cCorp}\swbox} & Corporate \& Commercial
  & 23{,}913 & 7.10
  & 31{,}462 & 12.22 \\
{\color{cGov}\swbox} & Government \& Official
  & 27{,}453 & 8.15
  & 32{,}013 & 12.43 \\
{\color{cNGO}\swbox} & NGO \& Think Tank
  & 14{,}520 & 4.31
  & 30{,}405 & 11.81 \\
{\color{cNews}\swbox} & News \& Journalism
  & 146{,}417 & 43.49
  & 79{,}600 & 30.91 \\
{\color{cOpin}\swbox} & Opinion \& Advocacy
  & 6{,}587 & 1.96
  & 6{,}983 & 2.71 \\
{\color{cRef}\swbox} & Reference \& Tertiary Source
  & 23{,}479 & 6.97
  & 34{,}468 & 13.39 \\
{\color{cUGC}\swbox} & User-Generated Content (UGC)
  & 4{,}554 & 1.35
  & 16{,}981 & 6.59 \\
\midrule
 & Total
  & 336{,}686 & 100.00
  & 257{,}486 & 100.00 \\
\bottomrule
\end{tabular}
\end{table}

\section{Media Domain Taxonomy Workflow}
\label{subsec:coding_manual}

\subsection*{Multi-Agent Workflow to Develop Coding Manual}

To avoid any ambiguity about the division of labour in what follows: the language models proposed candidate categories and later applied the finished manual, but they did not determine the scheme. The authors reconciled the four proposals, resolved the disagreements recorded in Table~\ref{tab:categoriess}, selected the final eight categories, and wrote the decision rules governing ambiguous cases. The internal consistency test reported below asks whether the resulting manual can be applied consistently; it is not a test of whether the categories are correct. Independent validation of the classifications against human coding is reported in the main text. We developed the Citation Content Coding Manual based on state-of-the-art multi-agent prompting workflow techniques involving the Deep Research models of ChatGPT5, Claude Sonnet 4.5, Gemini 2.5 Pro, and Grok4 (the same LLM on which Grokipedia is built)~\cite{hilbert2025ai}. 

\begin{figure}[htbp]
    \centering
    \includegraphics[width=0.8\textwidth]{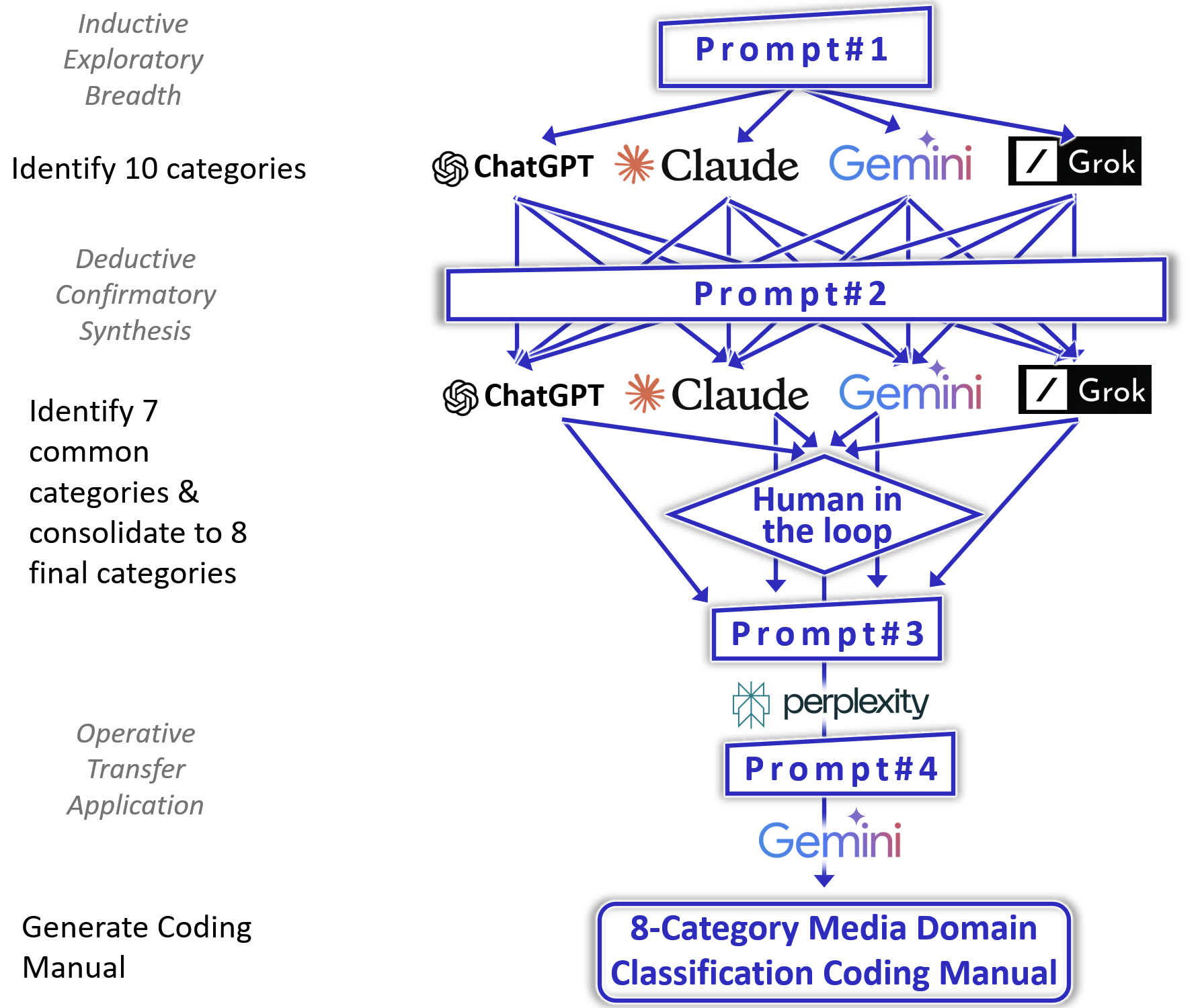}
    \caption{Multi-Agent Workflow to Develop Coding Manual}
    \label{fig:workflow}
\end{figure}

As shown in Figure \ref{fig:workflow}, we started with inductive and exploratory Deep Research (broad brainstorming, based on exemplary few-shot prompting), which identified the ten most common media domains. We then presented each of the four LLMs with the results from the other three LLMs and asked them to consolidate the results down to seven common categories. For Prompt\#1 and Prompt\#2, see ``Box: Prompts used in multi-agent workflow''.

\textbf{PROMPT\#1:} \\
You are an expert media analyst and librarian. Your task is to classify a list of domains based on their primary content. This first task consists in a deep research about different ways to classify media, for example: - 'Legacy News' (e.g., nytimes.com, wsj.com, bbc.com) - 'Digital News' (e.g., vox.com, dailywire.com, huffpost.com) - 'Opinion/Blog' (Primarily opinion, not reporting) - 'Academic/Journal' (e.g., nature.com, arxiv.org, jstor.org) - 'Corporate' (e.g., tesla.com, spacex.com, openai.com) - 'Government/NGO' (e.g., nasa.gov, un.org, brookings.edu) - 'Social Media' (e.g., x.com, reddit.com, facebook.com) - 'Reference' (e.g., britannica.com, dictionary.com) - 'Other/Unclear' (If it doesn't fit or is too obscure) - etc... Please do some deep research, focusing on academic sources, and come up with a final list of the most common 10 categories.

\medskip

\textbf{PROMPT\#2:} \\
Thanks. I asked three other experts to do the same. Please revise their answers carefully and consider if/how it can help to improve and complete your answer, making it more general and complete. Evaluate if you can merge some categories. If you can, great! Provide a final revised final list of the top seven most important media source categories.

\medskip

\textbf{PROMPT\#3:} \\
Based on this, create a final taxonomy, based on the eight categories outlined in the summary table, and make sure that they are defined in a way that is exhaustive and exclusive, means that any kind of media source can unequivocally be assigned to one, and only one, of the eight categories. Include some illustrative examples for each category. Then, imagine you are an expert in content coding and natural language processing. You are known to elaborate content coding manuals that are as clear and specific that they achieve a very high interrater reliability, such as evidenced by high Krippendorf alpha values. Please convert the previous conclusions of the 8-category taxonomy into a coding manual that allows content coders, especially in the form of Large Language Models (LLMs) to classify media sources into one of the eight categories. The input will be a simple link or platform name, and the LLM content coder should analyze/search that content provider and follow the instructions of the coding manual to clearly and unequivocally classify as belonging to one, and only one, of the eight categories. Provide a machine readable output document in the form of a content coding manual that can be used as a prompt for LLM content coders.

\medskip

\textbf{PROMPT\#4:} \\
Study the attached manual. Pivot to classification of citations instead of domains. How should the categories be changed? Wikipedia is full of self citations. For example, if citation is to a book but Wikipedia page of the book, it should still be categorized as an Academic \& Scholarly book.

\medskip
\hrule
\smallskip
\textbf{Sources (Archived Chat Histories):} \\
\footnotesize
ChatGPT: \url{https://chatgpt.com/share/69094d6b-58f4-8003-88d9-509113aeddd3} \\
Claude: \url{https://claude.ai/share/fc9862d0-6cf7-465d-8e4f-05d9eebea8f6} \\
Gemini: \url{https://gemini.google.com/share/ba003c0c7170} \\
Grok: \url{https://grok.com/share/c2hhcmQtNQ\%3D\%3D_e53d7382-6307-43c8-b1f5-1a690a688a98} \\
Perplexity: \url{https://www.perplexity.ai/search/attached-i-give-you-4-differen-Bh66BDLfR7ar.VLBWKmqpQ?preview=1#1}

\subsection*{Category Synthesis}

Reviewing the results by a human expert, we obtained Table \ref{tab:categoriess}, which shows that different LLMs decided to merge different categories, resulting in a total of eight different categories. We decided to keep all eight and asked Perplexity to develop an exclusive and exhaustive categorization, based on the four proposals, along the summary table logic, and then to develop a content coding manual that can be used for LLM-based classification (see Prompt\#3).

\begin{table}[htbp]
    \centering
    \caption{LLM Category Identification Summary}
    \label{tab:categoriess}
    \begin{tabular}{|l|c|c|c|c|}
        \hline
        \textbf{} & \textbf{ChatGPT5} & \textbf{Claude S.4.5} & \textbf{Gemini2.5} & \textbf{Grok4} \\
        \textbf{} & \textbf{Deep Research} & \textbf{Research} & \textbf{Pro Deep} & \textbf{Expert} \\
        \hline
        
        1 News \& Journalism & 1 & \multirow{2}{*}{1} & 1 & 1 \\ 
        \cline{1-2} \cline{4-5} 
        
        2 Opinion and Advocacy & \multirow{2}{*}{1} & & 1 & 1 \\ 
        \cline{1-1} \cline{3-5} 
        
        3 NGO, Civil Soc. \& Think Tank & & \multirow{2}{*}{1} & \multirow{2}{*}{1} & 1 \\ 
        \cline{1-2} \cline{5-5} 
        
        4 Government \& Official & 1 & & & 1 \\ 
        \hline
        
        5 Academic \& Scholarly & 1 & 1 & 1 & \multirow{2}{*}{1} \\ 
        \cline{1-4} 
        
        6 Reference \& Aggregation & 1 & 1 & 1 & \\ 
        \hline
        
        7 Corporate \& Commercial & 1 & 1 & 1 & 1 \\ 
        \hline
        
        8 Social Media \& User Content & 1 & 1 & 1 & 1 \\ 
        \hline
    \end{tabular}
\end{table}

\subsection*{Internal Validity Testing}

As an intermediate step, we tested the internal validity of the coding manual that resulted after prompt\#3, by feeding it as a prompt into ChatGPT5, Claude Sonnet 4.5, Gemini 2.5 Pro, and Grok4, together with the following instructions:

\begin{quote}
\textbf{Prompt\#3validity:}
You are an expert in content coding. Please study the attached 8-Category Taxonomy for Media Domain Classification: "LLM CONTENT CODING MANUAL".  Based on it, please research the 15 media domains attached and classify them into one of the eight categories described in the coding manual attached. Only chose one category from the taxonomy, the best fitting one. As output, provide a table with the media domains and their category names.
\end{quote}

The results, shown in Table \ref{tab:compressed_validity}, demonstrated high agreement. Krippendorff recommends relying on variables with $\alpha \geq 0.800$ and using values between 0.667 and 0.800 only for tentative conclusions~\cite{krippendorff2019content,hayes2007answering}. These values show that the four models apply the manual consistently. Because the same models proposed the candidate taxonomies, this is a consistency check on the manual rather than a validation of the classifications; the classifications themselves are validated against human coding in the main text (Table~\ref{tab:validation_agreement}).

Finally, prompt\#4 was used to revise the domain-focused manual so that it applied to individual citations rather than to whole domains. The two decision rules encoded in that revision, ``Decision rules for ambiguous citations, Rule 1'' and ``Decision rules for ambiguous citations, Rule 2'', were specified by the authors on the basis of the ambiguities identified during pilot coding, as described in the main text; the prompt served to propagate them consistently through the manual's wording and examples.

\begin{table}[htbp]
    \centering
    \scriptsize 
    \setlength{\tabcolsep}{1.2pt} 
    \caption{Intercoder Reliability Test Results}
    \label{tab:compressed_validity}

    \begin{minipage}[t]{0.32\textwidth}
        \centering
        \begin{tabular}{lcccc}
            \toprule
            \textbf{Domain} & \textbf{GP} & \textbf{Cl} & \textbf{Ge} & \textbf{Gr} \\
            \midrule
            digitalspy.co.uk & 1 & 1 & 1 & 1 \\
            estadao.com.br & 1 & 1 & 1 & 1 \\
            wagnerreese.com & 7 & 7 & 7 & 7 \\
            hapsc.org & 3 & 3 & 3 & 3 \\
            electionguide.org & 6 & 6 & 6 & 6 \\
            tasteiran.net & 7 & 7 & 7 & 7 \\
            capp.ca & 3 & 3 & 3 & 3 \\
            iba.edu.pk & 5 & 5 & 5 & 5 \\
            ballotpedia.org & 6 & 6 & 6 & 6 \\
            infobae.com & 1 & 1 & 1 & 1 \\
            c2es.org & 3 & 3 & 3 & 3 \\
            wayne.edu & 5 & 5 & 5 & 5 \\
            padidehzaban.ir & 7 & - & 5 & 7 \\
            cepa.org & 3 & 3 & 3 & 3 \\
            cell.com & 5 & 5 & 5 & 5 \\
            \midrule
            \textbf{K-Alpha} & \multicolumn{4}{c}{\textbf{0.982}} \\
            \bottomrule
        \end{tabular}
    \end{minipage}
    \hfill
    \begin{minipage}[t]{0.32\textwidth}
        \centering
        \begin{tabular}{lcccc}
            \toprule
            \textbf{Domain} & \textbf{GP} & \textbf{Cl} & \textbf{Ge} & \textbf{Gr} \\
            \midrule
            tarunias.com & 7 & 7 & 7 & 7 \\
            stanstedairport... & 7 & 7 & 7 & 7 \\
            georgetown.edu & 5 & 5 & 5 & 5 \\
            trustedreviews... & 1 & 1 & 1 & 1 \\
            bloodyelbow.com & 1 & 1 & 1 & 1 \\
            pagba.com & 3 & - & 3 & 3 \\
            manchester.ac.uk & 5 & 5 & 5 & 5 \\
            hitquarters.com & 6 & - & 1 & 7 \\
            thewrap.com & 1 & 1 & 1 & 1 \\
            heelbynature.com & 1 & - & 1 & 1 \\
            wdbj7.com & 1 & 1 & 1 & 1 \\
            onet.pl & 1 & 1 & 1 & 1 \\
            wisc.edu & 5 & 5 & 5 & 5 \\
            colonialwill... & 3 & 3 & 3 & 3 \\
            radaronline.com & 1 & - & 1 & 1 \\
            \midrule
            \textbf{K-Alpha} & \multicolumn{4}{c}{\textbf{0.904}} \\
            \bottomrule
        \end{tabular}
    \end{minipage}
    \hfill
    \begin{minipage}[t]{0.32\textwidth}
        \centering
        \begin{tabular}{lcccc}
            \toprule
            \textbf{Domain} & \textbf{GP} & \textbf{Cl} & \textbf{Ge} & \textbf{Gr} \\
            \midrule
            tracesofwar.com & 6 & - & 6 & 6 \\
            chelseafc.com & 7 & 7 & 7 & 7 \\
            tripshock.com & 7 & 7 & 7 & 7 \\
            southcoasttoday... & 1 & 1 & 1 & 1 \\
            livedesignonline... & 1 & 1 & 1 & 1 \\
            lionsrugby.com & 7 & 7 & 7 & 3 \\
            odi.org & 3 & 3 & 3 & 3 \\
            tradingeconomics... & 6 & 6 & 6 & 6 \\
            irishtimes.com & 1 & 1 & 1 & 1 \\
            essence.com & 1 & 1 & 1 & 1 \\
            chicagodetours... & 7 & 7 & 7 & 7 \\
            earlychicago.com & 6 & 6 & 6 & 6 \\
            discovermoose... & 1 & 1 & 1 & 1 \\
            icraa.org & 3 & 3 & 3 & 3 \\
            worldcargonews... & 1 & 1 & 1 & 1 \\
            \midrule
            \textbf{K-Alpha} & \multicolumn{4}{c}{\textbf{0.959}} \\
            \bottomrule
        \end{tabular}
    \end{minipage}

    \vspace{0.2cm}
    \footnotesize
    \textit{Note: GP=ChatGPT5, Cl=Claude S.4.5, Ge=Gemini 2.5 Pro, Gr=Grok4 Expert. Values 1-8 represent taxonomy categories. "-" indicates missing data.}
\end{table}

\subsection{Core Design Principles \& Decision rules for ambiguous citations}

\subsubsection{Two decision rules for ambiguous citations}
To handle the ambiguity between a host domain and a specific cited work, two rules must be applied \textit{before} the decision tree.

\begin{goldenrule}
\textbf{Rule 1: The Specific Work Principle} \\[1ex]
The classification of a \textit{specific cited work} (e.g., an op-ed, a journal article) overrules the general classification of its \textit{parent domain}.

\begin{itemize}[leftmargin=*, nosep]
    \item \textbf{Example:} A citation to an op-ed on \texttt{nytimes.com} is classified as \textbf{Opinion \& Advocacy}, even though the \texttt{nytimes.com} domain is broadly \textbf{News \& Journalism}.
\end{itemize}
\end{goldenrule}

\begin{goldenrule}
\textbf{Rule 2: The Look-Through Principle} \\[1ex]
If a citation links to a page that \textbf{hosts or describes} a specific work (like a Google Books entry, an Internet Archive scan, or a catalog page for a book or journal article), the citation must be classified as the \textbf{underlying work}, not the page that hosts or describes it.

\begin{itemize}[leftmargin=*, nosep]
    \item \textbf{Example:} The Wikipedia article on Buddhism cites Peter Harvey's \textit{An Introduction to Buddhism} (Cambridge University Press) through a link to its Google Books page. The citation is classified as \textbf{Academic \& Scholarly}, since the object invoked is the book, not Google Books.
    \item A citation that gives a book's author, title, page numbers, or ISBN but links the title to the book's Wikipedia page is classified by the book (e.g., \textbf{Academic \& Scholarly}), not as \textbf{Reference \& Tertiary Source}.
    \item A citation to the content of a Wikipedia article or another encyclopedia entry on a general topic remains \textbf{Reference \& Tertiary Source}.
\end{itemize}
\end{goldenrule}

\subsubsection{Core Principles}
\begin{enumerate}[nosep]
    \item \textbf{Sequential Decision Logic:} The classification algorithm uses a step-by-step decision tree that prioritizes the most distinct categories first.
    \item \textbf{Operationalized Criteria:} Each category is defined by the observable function of the \textit{work itself}.
    \item \textbf{Irrelevance Rules:} Classification must ignore:
        \begin{itemize}[nosep]
            \item \textbf{Content Topic:} A scientist's blog post (\textbf{UGC}) about politics is not \textbf{Government}. An op-ed (\textbf{Opinion}) about science is not \textbf{Academic}.
            \item \textbf{Business Model:} A paywalled news article is still \textbf{News}.
            \item \textbf{Format/Technology:} A news podcast is \textbf{News}. A PDF of a government report is \textbf{Government}.
        \end{itemize}
\end{enumerate}

\subsection{Quick Reference: Category Definitions}

\begin{table}[htbp]
\centering
\small
\begin{tabular}{@{}lp{4.5cm}p{4cm}p{3.5cm}@{}}
\toprule
\textbf{Category} & \textbf{Primary Function} & \textbf{Key Indicators (for the work)} & \textbf{Examples} \\
\midrule
\rowcolor{gray!10}
\textbf{Academic \& Scholarly} & Peer-reviewed, original research and scholarly communication. & Peer-review, university press, scholarly journal, dissertation, conference paper. & \texttt{nature.com} article, \texttt{jstor.org} article, academic book. \\

\textbf{Government \& Official} & Official information, data, services, or legal documents from government bodies. & Government agency author, court records, legislation, official statistics. & NASA report, Census data, \texttt{.gov} publication, UN resolution. \\

\rowcolor{gray!10}
\textbf{NGO \& Think Tank} & Mission-driven research and advocacy by non-profits. & Non-profit status, independent research, policy papers, advocacy reports. & Brookings report, Amnesty Int'l document, Greenpeace publication. \\

\textbf{News \& Journalism} & Original, factual reporting of current events. & Verifiable facts, journalistic standards, byline, reporting. & \texttt{nytimes.com} article, BBC report, Reuters dispatch. \\

\rowcolor{gray!10}
\textbf{Opinion \& Advocacy} & Persuasive commentary from an ideological perspective. & Op-ed, editorial, commentary, advocacy journalism, ideological positioning. & \texttt{huffpost.com} op-ed, \texttt{dailywire.com} article, blog post arguing a position. \\

\textbf{Corporate \& Commercial} & For-profit business products, services, and marketing. & Press release, annual report, company blog, product white paper, patent. & \texttt{apple.com} press release, Tesla annual report. \\

\rowcolor{gray!10}
\textbf{Reference \& Tertiary Source} & Organization and indexing of information from other sources (about a general topic). & Encyclopedia, dictionary, general topic summary. (See Rule 2). & \texttt{wikipedia.org} page for "Justice", \texttt{dictionary.com}. \\

\textbf{User-Generated Content (UGC)} & Informal content from a private individual. & User profile, forum, personal blog, non-professional social media post. & Tweet, Reddit comment, personal blog post, non-institutional YouTube video. \\
\bottomrule
\end{tabular}
\caption{Citation Category Definitions}
\label{tab:categories}
\end{table}

\subsection{Decision Tree for Citation Classification}

Follow these steps sequentially.

\begin{enumerate}[label=\textbf{Step \arabic*:}, wide, labelwidth=!, labelindent=0pt, nosep]
    \item \textbf{Is the \textit{work} peer-reviewed, original research (e.g., journal article, scholarly book)?}
        \begin{itemize}[nosep]
            \item \textbf{YES} $\rightarrow$ \textbf{ACADEMIC \& SCHOLARLY}
        \end{itemize}

    \item \textbf{NO. Is the \textit{work} an official publication from a government body (legislative, judicial, or executive)?}
        \begin{itemize}[nosep]
            \item \textbf{YES} $\rightarrow$ \textbf{GOVERNMENT \& OFFICIAL}
        \end{itemize}

    \item \textbf{NO. Is the \textit{work} a report or publication from a non-profit, mission-driven organization (e.g., think tank, NGO)?}
        \begin{itemize}[nosep]
            \item \textbf{YES} $\rightarrow$ \textbf{NGO, CIVIL SOCIETY \& THINK TANK}
        \end{itemize}
        
    \item \textbf{NO. Is the \textit{work} a specific, verifiable, factual report on current events by a journalist/publisher?}
        \begin{itemize}[nosep]
            \item \textbf{YES} $\rightarrow$ \textbf{NEWS \& JOURNALISM}
        \end{itemize}
        
    \item \textbf{NO. Is the \textit{work} a piece of commentary intended to persuade (e.g., op-ed, editorial, advocacy post)?}
        \begin{itemize}[nosep]
            \item \textbf{YES} $\rightarrow$ \textbf{OPINION \& ADVOCACY}
        \end{itemize}

    \item \textbf{NO. Is the \textit{work} a publication from a for-profit corporation about its business (e.g., press release, annual report)?}
        \begin{itemize}[nosep]
            \item \textbf{YES} $\rightarrow$ \textbf{CORPORATE \& COMMERCIAL}
        \end{itemize}

    \item \textbf{NO. Is the \textit{work} a tertiary source (e.g., encyclopedia, dictionary entry)?}
        \begin{itemize}[nosep]
            \item \textbf{YES} $\rightarrow$ \textbf{Apply "Look-Through Principle" (Rule 2):}
                \begin{itemize}[nosep]
                    \item \textit{Check:} Is this tertiary source \textit{about} a specific book, journal article, or other citable work?
                        \begin{itemize}[nosep] 
                            \item \textbf{YES:} \textbf{Go back to Step 1} and classify that \textit{underlying} work.
                            \item \textbf{NO:} Classify as \textbf{REFERENCE \& TERTIARY SOURCE}.
                        \end{itemize}
                \end{itemize}
        \end{itemize}
        
    \item \textbf{NO. Is the \textit{work} created by a private individual in an informal, non-institutional capacity (e.g., tweet, forum post)?}
        \begin{itemize}[nosep]
            \item \textbf{YES} $\rightarrow$ \textbf{USER-GENERATED CONTENT (UGC)}
        \end{itemize}
\end{enumerate}

\subsection{Boundary Cases with Decisions}

\begin{table}[htbp]
\centering
\begin{tabularx}{\linewidth}{@{}>{\raggedright\arraybackslash}X >{\centering\arraybackslash}m{2.4cm} >{\raggedright\arraybackslash}X@{}}
\toprule
\textbf{Case} & \textbf{Decision} & \textbf{Reasoning} \\
\midrule
Wall Street Journal factual article & \textbf{NEWS} & Original factual reporting (Rule~1). \\
Wall Street Journal op-ed & \textbf{OPINION} & Persuasive commentary (Rule~1). \\
Tesla blog post about technology & \textbf{CORPORATE} & Company-owned marketing/communication. \\
Wikipedia page for "History of Japan" & \textbf{REFERENCE} & Tertiary summary of a general topic. \\
Google Books page for Harvey's \textit{An Introduction to Buddhism} & \textbf{ACADEMIC} & ``Look-Through Principle'' (Rule~2) applied. \\
Kotkin's \textit{Stalin} with page numbers, title linked to its Wikipedia page & \textbf{ACADEMIC} & Classified by the book's bibliographic details (Rule~2). \\
Tweet from a journalist breaking news & \textbf{NEWS} & Factual reporting, despite platform. \\
Tweet from a journalist with an opinion & \textbf{OPINION} & Commentary, despite platform. \\
Tweet from a random user & \textbf{UGC} & Informal, non-institutional source. \\
Public radio (NPR, BBC) news report & \textbf{NEWS} & Journalism is primary function. \\
University press release & \textbf{CORPORATE} & Institutional promotion, not research. \\
University researcher's personal blog & \textbf{UGC} & Not institutionally published or peer-reviewed. \\
\bottomrule
\end{tabularx}
\caption{Citation Boundary Cases}
\end{table}

\subsection{Validation Checklist}
After classification, verify all conditions are met:
\begin{itemize}[label=\checkmark, nosep]
    \item Classification is based on the \textbf{specific work's function}.
    \item Classification can be traced through the decision tree.
    \item "Rules" 1 and 2 were considered and applied if necessary.
    \item Classification did NOT use content TOPIC as a criterion.
    \item Classification did NOT use business MODEL as a criterion.
    \item Classification did NOT use FORMAT/TECHNOLOGY as a criterion.
    \item Classification is MUTUALLY EXCLUSIVE (no other category equally valid).
\end{itemize}

\section{Data Collection}
\label{sec:appendix_data}

We analyzed two corpora of 1{,}000 matched article pairs each. The main corpus consists of the most-revised English Wikipedia articles, taken from Wikipedia's ranking of articles by number of revisions (Special:MostRevisions), retrieved through the MediaWiki API on 13 September 2026. To restrict the comparison to prose articles, we excluded titles beginning with ``List of'', ``Lists of'', ``Deaths in'', ``Timeline of'', ``Index of'', ``Outline of'' or ``Glossary of''; standalone years (e.g., 1991) and decades (e.g., 1990s); and discographies and records pages, which consist mainly of tables and whose citation density is not comparable with that of prose. Working down the ranking, we examined 1{,}048 titles: 20 had no Grokipedia page and 28 were excluded, leaving 1{,}000 pairs. The second corpus is drawn from Wikipedia's Vital Articles Level 3 list, retrieved on the same date (Appendix~\ref{sec:appendix_vital}).

Articles were matched by exact title, without fuzzy matching. Wikipedia pages were retrieved through the MediaWiki API (\url{https://en.wikipedia.org/w/api.php?action=parse&page=<Title>}) and Grokipedia pages from \url{https://grokipedia.com/page/<Title>}, with spaces replaced by underscores. A Wikipedia title had to be an article, not a redirect, missing page or disambiguation page. A Grokipedia page had to load without redirecting and without a ``does not exist'' placeholder. All pages were downloaded with the \texttt{requests} library (Python 3.10.12) and parsed with BeautifulSoup4 (4.14.3).

On Wikipedia, citations were read from every reference list on the page. Each footnote was split into the works it cites, and explanatory notes that cite no source were excluded. Citations generated by Wikipedia's citation templates were read directly. Shortened footnotes such as ``Kotkin 2014, p.~24'' were followed through their link to the full entry in the article's bibliography. Links to \texttt{wikipedia.org} were not collected, so citations without an external link, mostly books, were classified from their bibliographic text. On Grokipedia, each item of the article's reference list was recorded as one citation.

On Wikipedia, text was taken from the main content container and limited to headings, paragraphs and list items. Tables, figures, captions, formulas, reference lists and citation markers were removed first. On Grokipedia, text was taken from the main article container after removing the reference list, tables, figures, headers, footers, formulas and citation markers. On both platforms we also removed scripts and style elements.

\section{Classifier Performance by Category}
\label{sec:appendix_validation}

Table~\ref{tab:classification_examples} shows examples of classified citations from the most-revised corpus, one for each of several categories.

\begin{table}[htbp]
\centering
\small
\caption{\textbf{Examples of classified citations.} Real citations from the most-revised corpus with the category assigned by the classifier.}
\label{tab:classification_examples}
\renewcommand{\arraystretch}{1.2}
\begin{tabularx}{\textwidth}{@{} l l >{\raggedright\arraybackslash}X l @{}}
\toprule
\textbf{Platform} & \textbf{Article} & \textbf{Citation} & \textbf{Category} \\
\midrule
Wikipedia & Buddhism & Harvey (2013), \textit{An Introduction to Buddhism}, Cambridge University Press, linked via Google Books & Academic \& Scholarly \\
Wikipedia & Jimmy Carter & \textit{The Washington Post}, opinion section (\texttt{/opinions/}) & Opinion \& Advocacy \\
Grokipedia & 2016 U.S. presidential election & U.S. Census Bureau, ``Characteristics of Voters in the Presidential Election of 2016'' & Government \& Official \\
Grokipedia & 2003 invasion of Iraq & Brookings Institution, ``Iraq Without a Plan'' & NGO \& Think Tank \\
Grokipedia & 2003 invasion of Iraq & Britannica, ``Iraq War: Summary, Causes, Dates, Combatants, Casualties, \& Facts'' & Reference \& Tertiary \\
Grokipedia & 50 Cent & Reddit thread (r/todayilearned) & User-Generated Content \\
\bottomrule
\end{tabularx}
\end{table}

Table~\ref{tab:validation_percategory} reports precision, recall, and F1 of the production classifier against the two-coder consensus, computed on the 150 citations of the validation sample on which both coders agree (135 in Stratum A and 15 in Stratum B).

\begin{table}[htbp]
\centering\small
\caption[Classifier performance by category]{\textbf{Production classifier against the two-coder consensus, by category.} $n$ is the number of citations whose consensus label is that category; total $n=150$.}
\label{tab:validation_percategory}
\renewcommand{\arraystretch}{1.15}\setlength{\tabcolsep}{7pt}
\input{tables/tab_validation_percategory}
\end{table}

Table~\ref{tab:validation_errors} lists examples of disagreements. They cluster on adjacent pairs, above all \textit{Reference} versus \textit{User-Generated Content} and \textit{Opinion} versus \textit{NGO}, and no confusion type occurs more than three times in the sample.

\begin{table}[htbp]
\centering\scriptsize
\caption[Classifier disagreements]{\textbf{Disagreements between the classifier and the human consensus.}}
\label{tab:validation_errors}
\renewcommand{\arraystretch}{1.12}
\input{tables/tab_validation_errors}
\end{table}

\section{Model Fit for Citation Scaling}
\label{sec:appendix_scaling}

Table~\ref{tab:scaling_model_fit} reports the regression statistics behind Figure~\ref{fig:citations_vs_words}. For each platform we fitted a linear and a quadratic OLS model of citation count on word count and retained the quadratic term only where it was significant at $\alpha=0.05$. The table also reports the log--log elasticity.

\begin{table}[htbp]
\centering\small
\caption[Regression statistics for citation scaling]{\textbf{Citation count as a function of article length.} OLS, $N=1{,}000$ articles per platform.}
\label{tab:scaling_model_fit}
\renewcommand{\arraystretch}{1.2}
\setlength{\tabcolsep}{8pt}
\input{tables/tab_scaling_model_fit}
\end{table}

\section{Most- and Least-Cited Articles}
\label{sec:appendix_topbottom}

\begin{figure}[htbp]
    \centering
    \includegraphics[width=\linewidth]{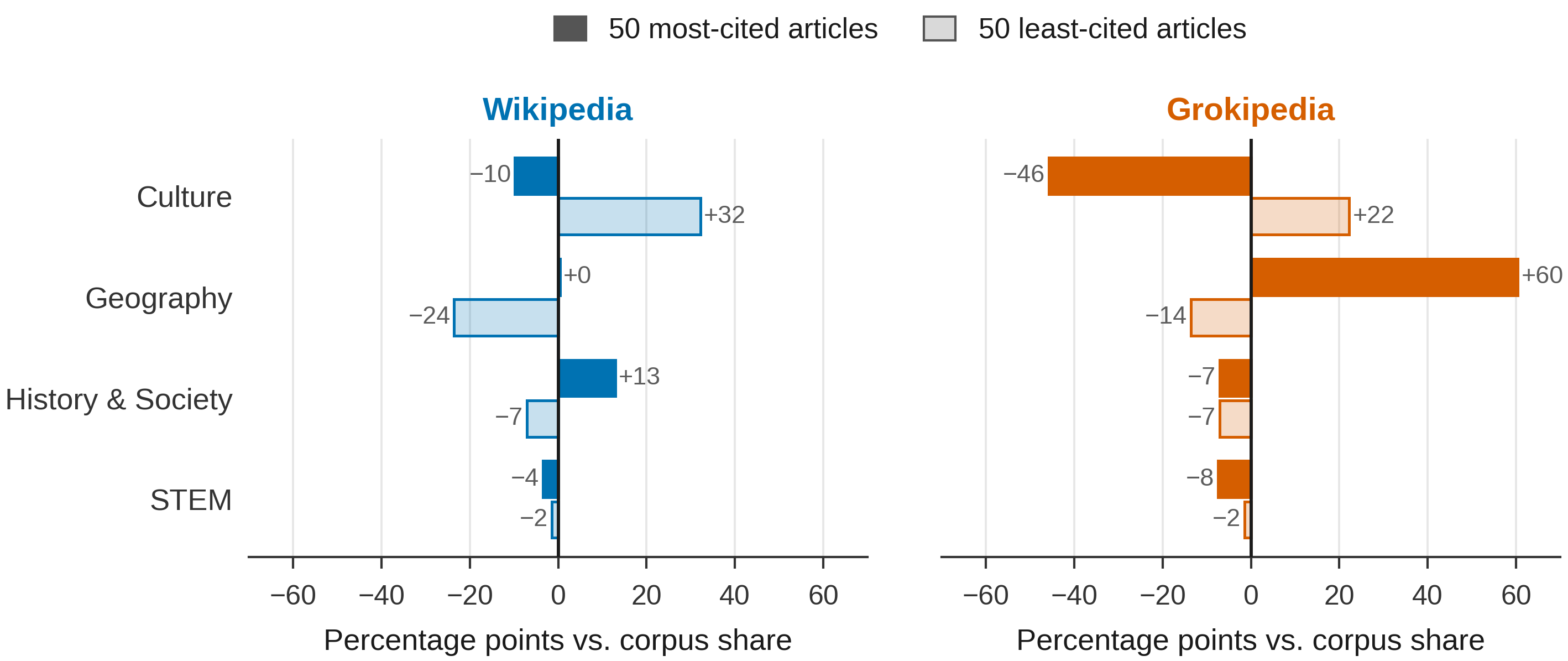}
    \caption[Topics of the most- and least-cited articles]{\textbf{Topics of the most- and least-cited articles.} For the 50 articles with the most citations (solid) and the fewest (pale) on each platform, the share of each topic domain minus that domain's share of the full corpus, in percentage points. Zero means the topic appears as often as in the corpus as a whole.}
    \label{fig:topbottom}
\end{figure}

Figure~\ref{fig:topbottom} shows the topics of the 50 most-cited and 50 least-cited articles on each platform. On Wikipedia, the 50 most-cited articles are spread across \textit{Culture}, \textit{Geography}, and \textit{History \& Society}, and are led by political biographies (Vladimir Lenin, Joseph Stalin, Nelson Mandela, Muammar Gaddafi). On Grokipedia, the 50 most-cited articles are almost all \textit{Geography}, mostly countries and U.S. states (History of the United States, India, Brazil, the United States), with Elon Musk the main exception. On both platforms the 50 least-cited articles are mostly \textit{Culture}: television series, video games, athletes, and entertainers.

Table~\ref{tab:martintable} gives examples. Wikipedia's policy on biographies of living persons, which requires that contentious material be sourced inline and that contentious material that is unsourced or poorly sourced be removed immediately~\cite{wikipedia_BLP}, is one candidate explanation for the concentration of Wikipedia's citations on biographies, though our design cannot confirm it.

\begin{table}[htbp]
\centering\small
\caption[Most- and least-cited articles]{\textbf{Most- and least-cited articles on each platform.} Examples from the top and bottom of each platform's ranking by number of citations.}
\label{tab:martintable}
\renewcommand{\arraystretch}{1.12}
\setlength{\tabcolsep}{6pt}
\begin{tabular}{@{} l r @{\hspace{2.2em}} l r @{}}
\toprule
\multicolumn{2}{c}{\textbf{Wikipedia}} & \multicolumn{2}{c}{\textbf{Grokipedia}} \\
\cmidrule(r{1.2em}){1-2}\cmidrule(l){3-4}
Article & Citations & Article & Citations \\
\midrule
\multicolumn{4}{@{}l}{\textit{Most cited}} \\
\addlinespace[2pt]
Vladimir Lenin   & 1{,}533 & History of the United States & 914 \\
Joseph Stalin    & 1{,}371 & Elon Musk                    & 832 \\
Nelson Mandela   & 1{,}188 & India                        & 686 \\
Muammar Gaddafi  &     945 & Brazil                       & 661 \\
Fidel Castro     &     911 & United States                & 645 \\
\midrule
\multicolumn{4}{@{}l}{\textit{Least cited}} \\
\addlinespace[2pt]
Kenz\=o Shirai       & 29 & Hyderabad             & 23 \\
Total Drama Island   & 34 & Kareena Kapoor        & 48 \\
Drake \& Josh        & 49 & Shadow the Hedgehog   & 63 \\
High School Musical  & 53 & The Voice (franchise) & 72 \\
Cartoon Network      & 59 & WWE Championship      & 85 \\
\bottomrule
\end{tabular}
\end{table}

\section{Sensitivity Analysis of Network Structures}
\label{sec:appendix_sensitivity}

We conducted an analysis that is more sensitive to rare epistemological profiles—categories that appear infrequently but nonetheless provide distinctive signatures for specific subsets of articles. In the raw co-occurrence networks (Figure~\ref{fig:co_occurrence_nets}), such low-frequency profiles often “fall through the cracks” because the visualization is dominated by high-volume categories.

\begin{figure}[htbp]
    \centering
    \begin{subfigure}[b]{0.48\textwidth}
        \centering
        \includegraphics[width=\textwidth]{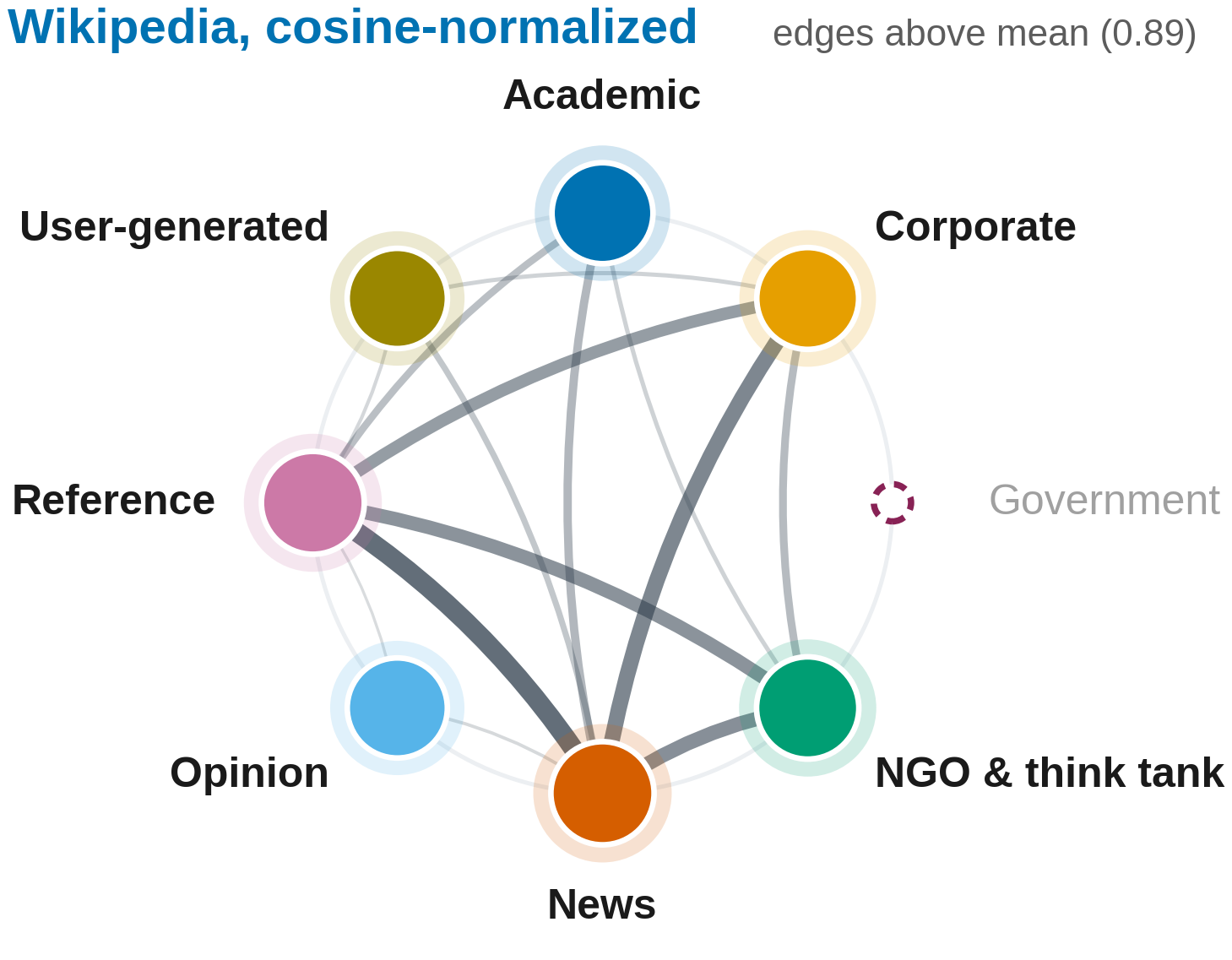}
        \caption{Wikipedia (Normalized)}
        \label{fig:sens_wiki}
    \end{subfigure}
    \hfill
    \begin{subfigure}[b]{0.48\textwidth}
        \centering
        \includegraphics[width=\textwidth]{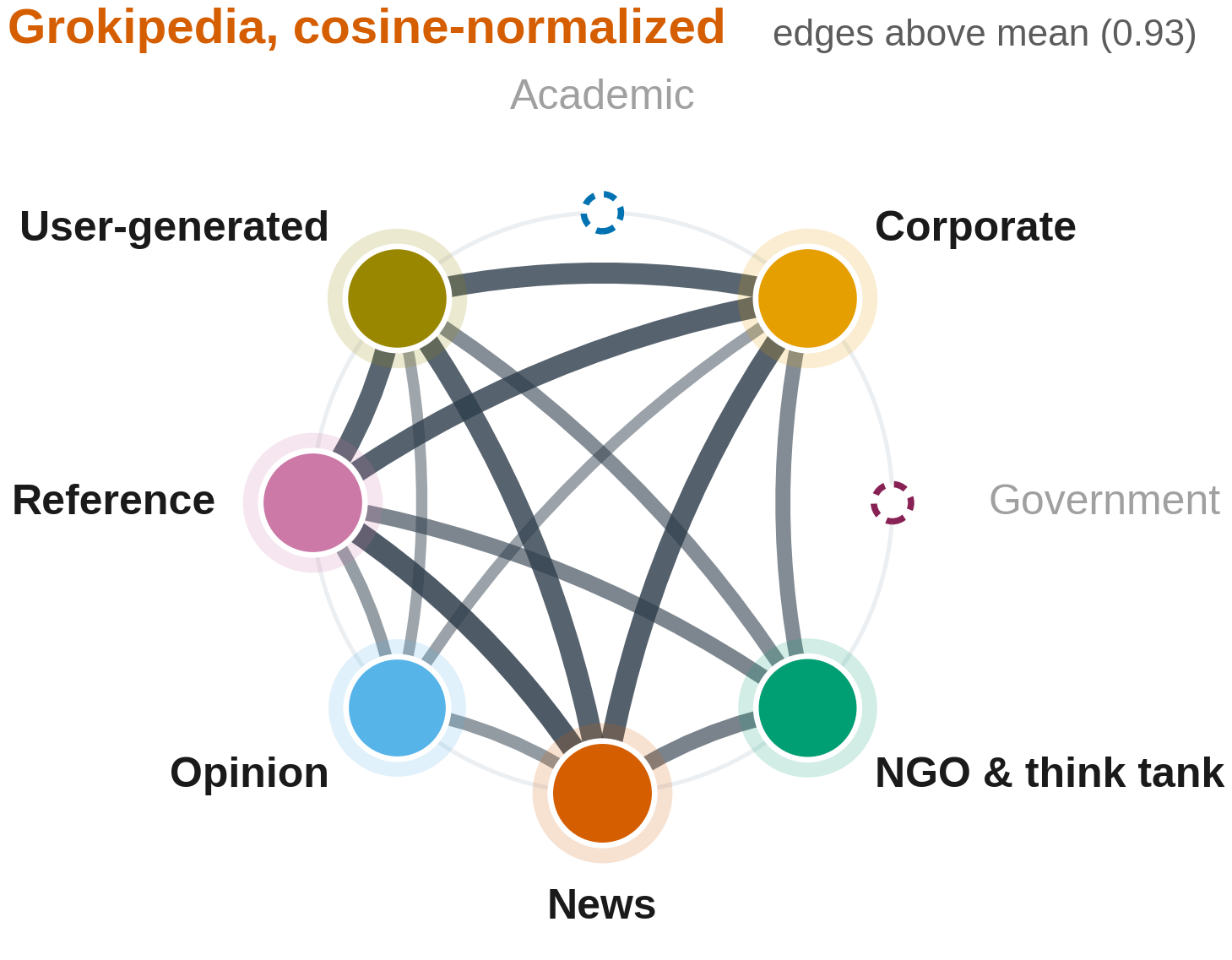}
        \caption{Grokipedia (Normalized)}
        \label{fig:sens_grok}
    \end{subfigure}
    \caption{\textbf{Sensitivity Analysis of Epistemic Backbones (Cosine Similarity).} Edges represent the cosine similarity between citation categories calculated from per-article presence vectors, retaining edges above each platform's mean similarity. Hollow dashed nodes mark categories with no edge above the cutoff.}
    \label{fig:sensitivity_networks}
\end{figure}

To address this, we applied Cosine Similarity normalization to examine the backbone of category co-occurrences while controlling for category frequency. For each article, we constructed a binary vector indicating the presence or absence of the eight epistemic categories. After transposing the Article–Category matrix, we computed the Cosine Similarity between all category pairs, yielding a category–category similarity matrix. This normalization reduces the influence of highly frequent categories and allows low-frequency categories to form connections if they consistently co-occur with others. We then filtered the networks to retain only edges with similarity scores strictly greater than the global mean. The resulting networks, presented in Figure~\ref{fig:sensitivity_networks}, offer a complementary perspective to the raw co-occurrence graphs by highlighting structurally significant relationships independent of raw volume.

The cosine-normalized backbones (Figure~\ref{fig:sensitivity_networks}) retain the same nodes as the raw co-occurrence networks: \textit{Academic \& Scholarly} is part of Wikipedia's core and drops out of Grokipedia's, and \textit{Government \& Official} falls below the cutoff on both platforms. The main result of the co-occurrence analysis is therefore not an artifact of category frequency, pushing these connections above the filtering threshold.

\subsection*{Robustness of the Co-occurrence Cutoff}
\label{sec:appendix_cooc_threshold}

The co-occurrence backbones in Figure~\ref{fig:co_occurrence_nets} retain edges whose weight exceeds the platform's mean edge weight. Because any such cutoff is arbitrary, Table~\ref{tab:cooc_threshold} repeats the construction under four rules: the mean, the median, and the 60th and 75th percentiles of the within-platform edge-weight distribution. The result that motivates the analysis is unchanged under all four: \textit{Academic \& Scholarly} belongs to Wikipedia's core and falls below the cutoff in Grokipedia's, and \textit{Government \& Official} falls below the cutoff on both platforms. The one rule-dependent node is \textit{User-Generated Content}, which leaves Wikipedia's core at the 75th percentile while remaining in Grokipedia's, so the contrast in UGC's structural position is if anything sharper under the stricter rule.

\begin{table}[htbp]
\centering\small
\caption[Robustness of the co-occurrence cutoff]{\textbf{Co-occurrence backbones under four cutoff rules.} \textit{Value}: the edge-weight cutoff implied by each rule, computed within platform. \textit{Edges}: edges retained out of 28 possible pairs.}
\label{tab:cooc_threshold}
\renewcommand{\arraystretch}{1.15}\setlength{\tabcolsep}{7pt}
\input{tables/tab_cooc_threshold}
\end{table}

\subsection*{Statistical Effect Sizes of Epistemic Diversity}
\label{sec:appendix_Effect Sizes}

To quantify the magnitude of the differences in epistemic diversity, we calculated paired Cohen's $d_z$ effect sizes and 95\% confidence intervals for each topic. Effect sizes are paired ($d_z$ = mean difference divided by the standard deviation of the differences). The differences are not normally distributed (Table~\ref{tab:entropy_robustness}), so we report the Wilcoxon signed-rank test and the matched-pairs rank-biserial correlation $r_{rb}$ as distribution-free checks. No transformation was applied, since entropy is bounded and on the same scale for both platforms. As detailed in Table \ref{tab:entropy_stats}, effect sizes are large in all four domains, from $d_z=0.88$ for \textit{STEM} to $1.25$ for \textit{History \& Society}, and Grokipedia's entropy is higher in 84\% to 93\% of pairs within each domain.

\begin{table}[htbp]
\centering
\small
\caption[Paired comparisons of source-mix entropy]{\textbf{Paired comparisons of source-mix entropy across topic groups.} Differences are Grokipedia $-$ Wikipedia. \textit{Distribution}: skewness and Shapiro--Wilk $p$ of the paired differences. \textit{Effect size}: mean difference (bits) and paired Cohen's $d_z$. \textit{Rank-based}: matched-pairs rank-biserial correlation $r_{rb}$ and the share of pairs in which Grokipedia's entropy is higher. Wilcoxon signed-rank $p<0.001$ in every row.}
\label{tab:entropy_stats}\label{tab:entropy_robustness}
\renewcommand{\arraystretch}{1.15}
\setlength{\tabcolsep}{6pt}
\begin{tabular}{@{} l r rr rr rr @{}}
\toprule
 & & \multicolumn{2}{c}{\textbf{Distribution}} & \multicolumn{2}{c}{\textbf{Effect size}} & \multicolumn{2}{c}{\textbf{Rank-based}} \\
\cmidrule(lr){3-4}\cmidrule(lr){5-6}\cmidrule(l){7-8}
\textbf{Topic} & $N$ & Skew & Shapiro $p$ & Mean diff. & $d_z$ & $r_{rb}$ & \% higher \\
\midrule
Culture            & 517 & 0.73 & $<$0.001 & $+$0.524 & 1.16 & 0.93 & 89.9 \\
Geography          & 296 & 1.49 & $<$0.001 & $+$0.505 & 1.00 & 0.96 & 92.9 \\
History \& Society & 114 & 0.86 & $<$0.001 & $+$0.741 & 1.25 & 0.97 & 93.0 \\
STEM               &  73 & 0.84 & 0.019    & $+$0.421 & 0.88 & 0.83 & 83.6 \\
\midrule
All                & 1{,}000 & 1.06 & $<$0.001 & $+$0.535 & 1.09 & 0.94 & 90.7 \\
\bottomrule
\end{tabular}
\end{table}

\subsection*{Robustness of Homophily Networks}
\label{sec:appendix_e}

To address potential sensitivity to the cosine similarity threshold, we calculated the Assortativity Coefficient ($r$) across a range of thresholds. Table \ref{tab:assortativity_sensitivity} confirms that Grokipedia consistently exhibits higher assortativity than Wikipedia at every threshold tested, with Grokipedia's coefficient 1.7 to 2.4 times Wikipedia's.

\begin{table}[htbp]
\centering
\small
\caption[Sensitivity of the assortativity coefficient]{\textbf{Sensitivity of the topic assortativity coefficient $r$ to the cosine threshold.} \textit{Edges}: edges retained at each threshold; $N=1{,}000$ articles per platform.}
\label{tab:assortativity_sensitivity}
\renewcommand{\arraystretch}{1.18}
\setlength{\tabcolsep}{12pt}
\input{tables/tab_assortativity_sensitivity}
\end{table}

\subsection{Topic Mixing Matrices}
\label{sec:appendix_mixing}

\begin{figure}[htbp]
  \centering
  \begin{minipage}[b]{0.48\textwidth}
    \centering
    \includegraphics[width=\textwidth]{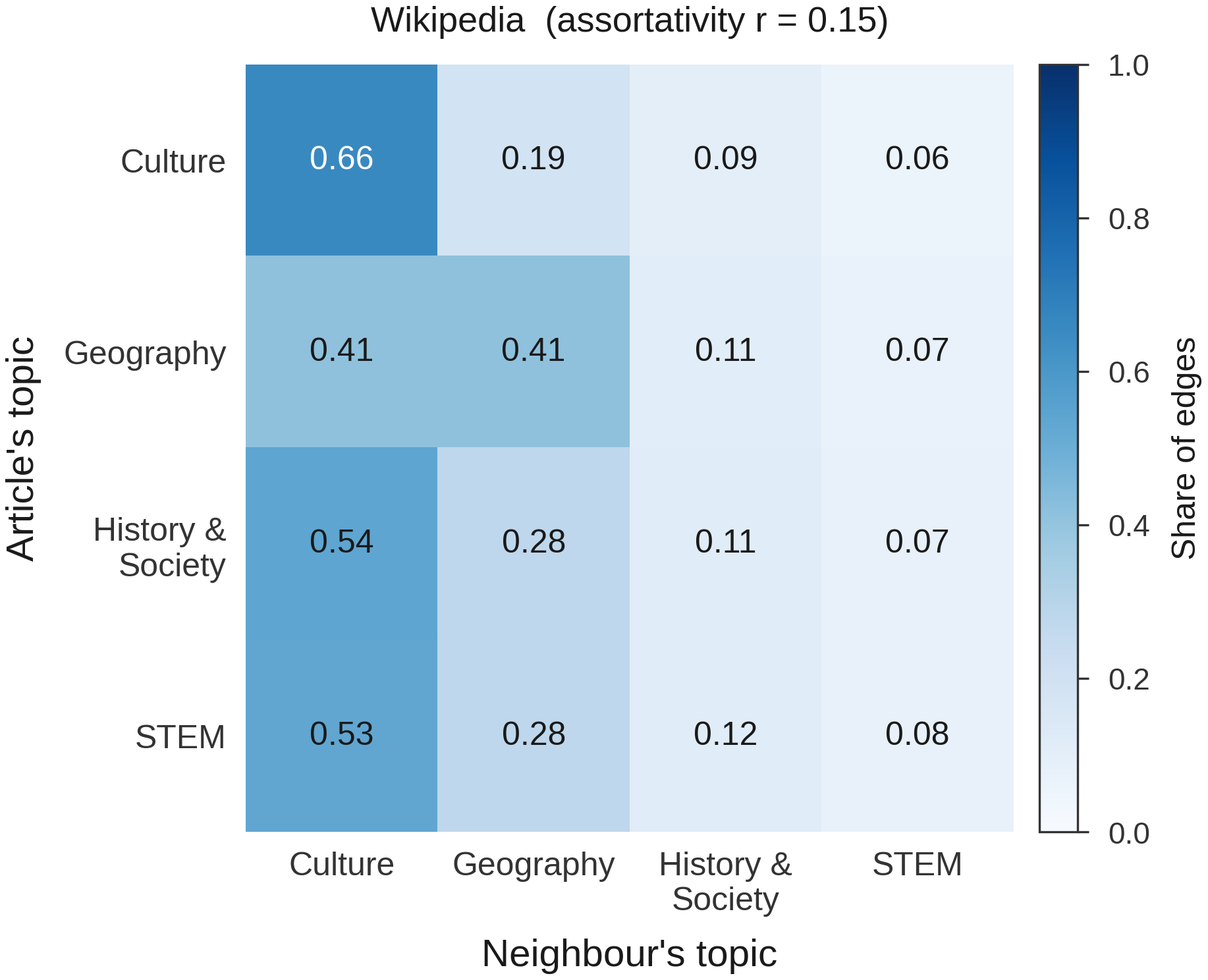}
    \caption*{(a) Wikipedia ($r = 0.15$)}
  \end{minipage}
  \hfill
  \begin{minipage}[b]{0.48\textwidth}
    \centering
    \includegraphics[width=\textwidth]{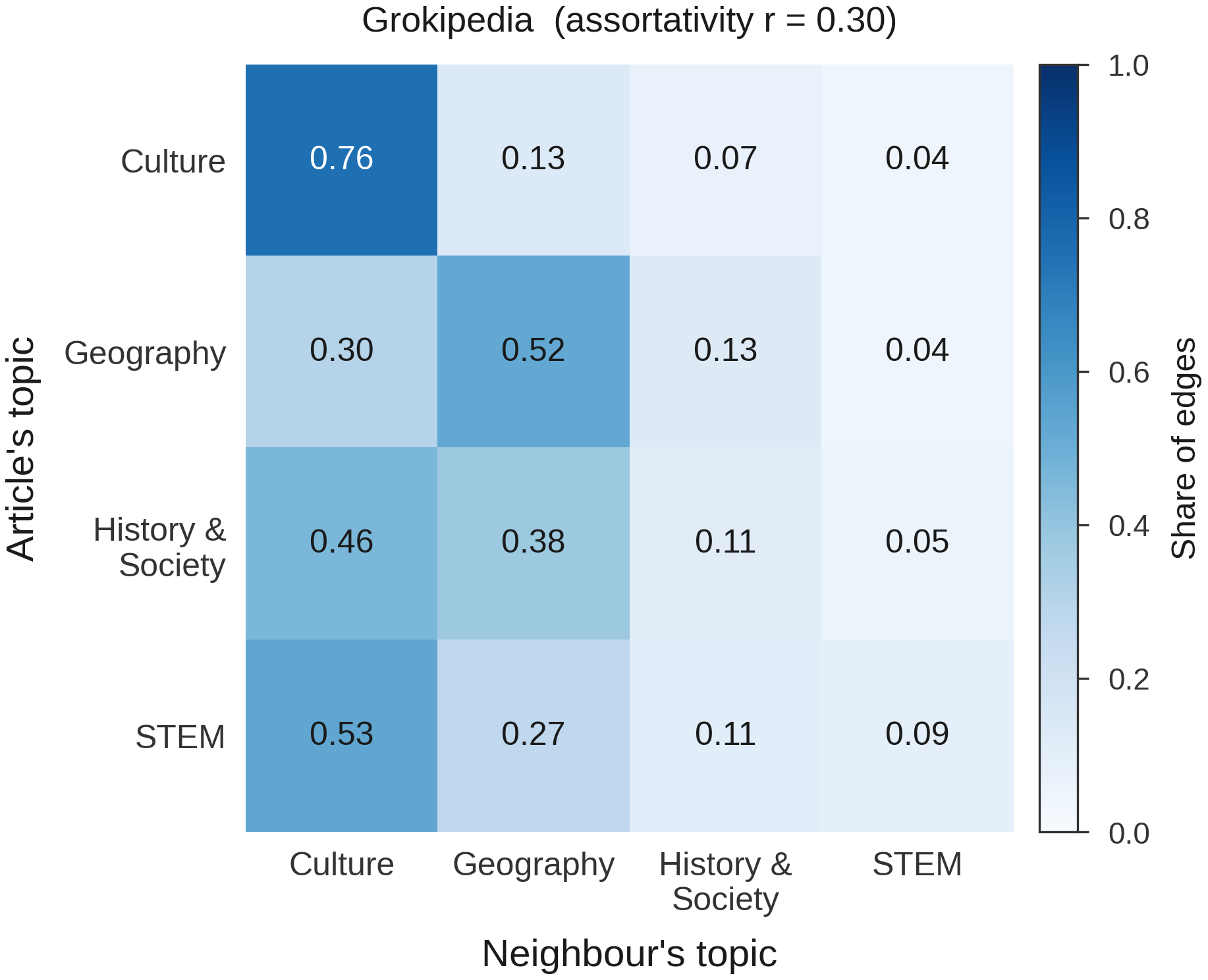}
    \caption*{(b) Grokipedia ($r = 0.30$)}
  \end{minipage}
   
  \caption{\textbf{Topic Mixing Matrices.} 
  Heatmaps display the row-normalized mixing matrix, representing the conditional probability that an article of a given topic connects to a target topic based on sourcing similarity ($>0.75$). Grokipedia has higher diagonal shares for \textit{Culture} (0.76 versus 0.66) and \textit{Geography} (0.52 versus 0.41).}
  \label{fig:mixing_matrices}
\end{figure}

To further examine the observed differences in topic assortativity, we calculated row-normalized mixing matrices for the article similarity networks (Figure~\ref{fig:mixing_matrices}). Each cell represents the proportion of existing connections from articles in one topic domain to articles in another, using a cosine similarity threshold of 0.75.

Grokipedia exhibits higher within-topic connectivity for \textit{Culture} (76\% versus 66\%) and \textit{Geography} (52\% versus 41\%). In Wikipedia, \textit{Geography} articles connect equally to \textit{Culture} and \textit{Geography} (41\% each), whereas in Grokipedia, 52\% of their connections remain within \textit{Geography} and 31\% connect to \textit{Culture}. In both platforms, connections from \textit{History \& Society} and \textit{STEM} are predominantly directed toward the two larger domains. These patterns suggest that Grokipedia's higher assortativity is driven mainly by stronger separation between cultural and geographic articles. Because mixing proportions are influenced by topic sizes, we interpret these results alongside the assortativity coefficients reported in the main text.

\section{Generalization to Wikipedia's Vital Articles}
\label{sec:appendix_vital}

To test whether our main results depend on selecting articles by editing activity, we repeated the analysis of epistemic profiles and citation density on a second corpus selected by importance: Wikipedia's Vital Articles (Level 3), a community-curated list of core topics, matched to 200 Grokipedia articles with the same title and collected in September 2026. Table~\ref{tab:vital_generalization} compares the two corpora. Every category moves in the same direction in both: Grokipedia cites proportionally fewer \textit{Academic \& Scholarly} and \textit{News \& Journalism} sources and more of each of the other six categories. Grokipedia's academic share is also lower in most matched pairs (69\% in the most-revised corpus and 88\% in the Vital corpus; Wilcoxon signed-rank $p<0.001$ in both). In both corpora, Grokipedia cites fewer sources per 1000 words than Wikipedia, and its citation density varies much less across articles. Absolute shares differ between the two corpora, which cover different topics, but the direction of the differences is the same.

\begin{table}[htbp]
\centering
\small
\caption{\textbf{Epistemic profile and citation density in the most-revised and Vital Articles corpora.} Shares are percentages of each platform's citations. Citation density is citations per 1{,}000 words.}
\label{tab:vital_generalization}
\renewcommand{\arraystretch}{1.2}
\setlength{\tabcolsep}{6pt}
\input{tables/tab_vital_generalization}
\end{table}

\end{appendices}

\end{document}

%% file: tables/tab_grok_time.tex
\begin{tabular}{@{} l r rr rr @{}}
\toprule
& & \multicolumn{2}{c}{\textbf{Median}} & \multicolumn{2}{c}{\textbf{Change}} \\
\cmidrule(lr){3-4}\cmidrule(l){5-6}
\textbf{Measure} & $n$ & Nov 2025 & Sep 2026 & Median \% & $p$ \\
\midrule
Words & 69 & 14{,}123 & 13{,}326 & +0.7 & 0.959 \\
Citations & 69 & 306 & 290 & +2.0 & 0.038 \\
Citations per 1k words & 69 & 22.3 & 21.9 & +0.1 & 0.576 \\
\midrule
\multicolumn{6}{@{}l}{\textit{Source profile}} \\
\addlinespace[2pt]
\multicolumn{2}{@{}l}{Jensen--Shannon distance, Nov $\rightarrow$ Sep, median [IQR]} & \multicolumn{4}{r}{0.101 [0.063, 0.159]} \\
\multicolumn{2}{@{}l}{Composition shifted, sum of $|\Delta|$} & \multicolumn{4}{r}{9.0 points} \\
\bottomrule
\end{tabular}

%% file: tables/tab_validation_percategory.tex
\begin{tabular}{l c ccc}
\toprule
\textbf{Category} & $n$ & \textbf{Precision} & \textbf{Recall} & \textbf{F1} \\
\midrule
Academic \& Scholarly & 17 & 0.94 & 0.94 & 0.94 \\
Corporate \& Commercial & 21 & 0.95 & 0.90 & 0.93 \\
Government \& Official & 20 & 0.87 & 1.00 & 0.93 \\
NGO \& Think Tank & 23 & 0.95 & 0.83 & 0.88 \\
News \& Journalism & 20 & 1.00 & 0.95 & 0.97 \\
Opinion \& Advocacy & 12 & 0.85 & 0.92 & 0.88 \\
Reference \& Tertiary & 15 & 0.79 & 1.00 & 0.88 \\
User-Generated Content & 22 & 0.95 & 0.82 & 0.88 \\
\midrule
Total & 150 & & & \\
\bottomrule
\end{tabular}

%% file: tables/tab_validation_errors.tex
\begin{tabular}{p{2.6cm} p{2.6cm} c p{7.2cm}}
\toprule
\textbf{Classifier} & \textbf{Human consensus} & $n$ & \textbf{Example citation} \\
\midrule
Reference \& Tertiary & User-Generated Content & 3 & \textit{"Communism" . Thinkquest . Archived from the original on 9 April 2000 . Retrieved 29 January 2012 .} \\
Opinion \& Advocacy & NGO \& Think Tank & 2 & \textit{Uniquely Bad—But Not Uniquely American - City Journal} \\
User-Generated Content & Corporate \& Commercial & 1 & \textit{"Floyd Estel Parton" . Archived from the original on August 2, 2025 . Retrieved August 15, 2025 .} \\
Reference \& Tertiary & Academic \& Scholarly & 1 & \textit{"Storia del teatro: lo spazio scenico in Toscana" (in Italian) . Retrieved 28 July 2022 .} \\
Corporate \& Commercial & News \& Journalism & 1 & \textit{"World Air Forces 2025" . Flightglobal : 33. Archived from the original on 21 January 2025 . Retrieved 5 March} \\
\bottomrule
\end{tabular}

%% file: tables/tab_scaling_model_fit.tex
\begin{tabular}{lcc}
\toprule
& \textbf{Wikipedia} & \textbf{Grokipedia} \\
\midrule
\multicolumn{3}{@{}l}{\textit{Linear model}} \\
Intercept & 64.30 (12.95) & 7.29 (9.04) \\
Slope (per 1{,}000 words) & 26.70 (1.41) & 21.23 (0.84) \\
Adjusted $R^2$ & 0.37 & 0.66 \\
\addlinespace
\multicolumn{3}{@{}l}{\textit{Quadratic model}} \\
Intercept & $-$73.89 (24.16) & $-$36.56 (18.81) \\
Linear term (per 1{,}000 words) & 57.52 (5.36) & 27.84 (3.13) \\
Quadratic term (per $(1{,}000\text{ words})^2$) & $-$1.48 (0.27) & $-$0.22 (0.12) \\
$p$, quadratic term & < 0.001 & 0.070 \\
Adjusted $R^2$ & 0.40 & 0.67 \\
\addlinespace
Model retained ($\alpha=0.05$) & quadratic & linear \\
Elasticity $b$ [95\% CI], log--log & 0.89 [0.82, 0.96] & 1.02 [0.97, 1.08] \\
\bottomrule
\end{tabular}

%% file: tables/tab_cooc_threshold.tex
\begin{tabular}{@{} l rrc rrc @{}}
\toprule
& \multicolumn{3}{c}{\textbf{Wikipedia}} & \multicolumn{3}{c}{\textbf{Grokipedia}} \\
\cmidrule(lr){2-4}\cmidrule(l){5-7}
\textbf{Cutoff rule} & Value & Edges & Academic in core & Value & Edges & Academic in core \\
\midrule
Mean & 783 & 14 & yes & 854 & 14 & no \\
Median & 782 & 14 & yes & 858 & 14 & no \\
60th pct. & 797 & 11 & yes & 891 & 11 & no \\
75th pct. & 836 & 7 & yes & 927 & 7 & no \\
\bottomrule
\end{tabular}

%% file: tables/tab_assortativity_sensitivity.tex
\begin{tabular}{ccccc}
\toprule
\textbf{Threshold} & \textbf{Wikipedia $r$} & \textbf{Edges} & \textbf{Grokipedia $r$} & \textbf{Edges} \\
\midrule
0.60 & 0.087 & 299{,}040 & 0.150 & 292{,}961 \\
0.65 & 0.105 & 278{,}021 & 0.196 & 261{,}025 \\
0.70 & 0.127 & 255{,}287 & 0.245 & 228{,}333 \\
0.75 & 0.152 & 230{,}423 & 0.298 & 195{,}128 \\
0.80 & 0.179 & 201{,}505 & 0.354 & 160{,}241 \\
0.85 & 0.200 & 170{,}078 & 0.406 & 124{,}032 \\
0.90 & 0.212 & 133{,}765 & 0.444 & 84{,}327 \\
0.95 & 0.202 & 83{,}436 & 0.478 & 38{,}471 \\
\bottomrule
\end{tabular}

%% file: tables/tab_vital_generalization.tex
\begin{tabular}{@{} l rr rr @{}}
\toprule
 & \multicolumn{2}{c}{\textbf{Most-revised}} & \multicolumn{2}{c}{\textbf{Vital Articles}} \\
\cmidrule(lr){2-3}\cmidrule(l){4-5}
 & Wikipedia & Grokipedia & Wikipedia & Grokipedia \\
\midrule
Matched article pairs & \multicolumn{2}{c}{1,000} & \multicolumn{2}{c}{200} \\
\addlinespace
\multicolumn{5}{@{}l}{\textit{Epistemic profile (\%)}} \\
\quad Academic \& Scholarly & 26.7 & 9.9 & 61.8 & 35.8 \\
\quad Government \& Official & 8.2 & 12.4 & 8.7 & 13.1 \\
\quad News \& Journalism & 43.5 & 30.9 & 11.4 & 8.9 \\
\quad NGO \& Think Tank & 4.3 & 11.8 & 4.0 & 11.7 \\
\quad Corporate \& Commercial & 7.1 & 12.2 & 1.8 & 9.4 \\
\quad Opinion \& Advocacy & 2.0 & 2.7 & 1.1 & 1.5 \\
\quad Reference \& Tertiary & 7.0 & 13.4 & 10.4 & 15.6 \\
\quad User-Generated Content & 1.4 & 6.6 & 0.8 & 3.9 \\
\addlinespace
Citation density, mean (SD) & 34.0 (16.0) & 21.8 (5.5) & 24.0 (12.9) & 20.5 (5.2) \\
\bottomrule
\end{tabular}